\documentclass{article}

\usepackage{tabularx}
\usepackage[page]{appendix}
\usepackage{xpatch}
\xpretocmd{\appendixpagename}{\sffamily}{}{}
\usepackage[utf8]{inputenc} 
\usepackage[T1]{fontenc}    
\usepackage{hyperref}      
\usepackage{booktabs}     
\usepackage{amsfonts}       
\usepackage{nicefrac}       
\usepackage{microtype}      
\usepackage{lipsum}		
\usepackage{graphicx}
\usepackage{doi}
\usepackage{amsmath}
\usepackage{amssymb}
\usepackage{multirow}
\usepackage[a4paper]{geometry}
\usepackage{float}
\usepackage{subfigure}
\usepackage{booktabs}
\usepackage{verbatim}
\usepackage[inline]{enumitem}
\usepackage{adjustbox}
\usepackage{graphicx}
\usepackage[dvipsnames]{xcolor}
\usepackage{natbib}
\usepackage{cite}
\usepackage{xr-hyper}
\usepackage{hyperref}
\providecommand{\keywords}[1]{\textbf{\textit{Keywords:}} #1}

\title{Impact of Jittering on Raster- and Distance-based Geostatistical Analyses of DHS Data}

\date{}

\author{Umut Altay\qquad John Paige\qquad Andrea Riebler\qquad \\Geir-Arne Fuglstad\\ \\Department of Mathematical Sciences, Norwegian University \\of Science and Technology, Trondheim, Norway}

\begin{document}
\maketitle
\begin{abstract}
Fine-scale covariate rasters are routinely used in geostatistical models for mapping demographic and health indicators based on household surveys from the Demographic and Health Surveys (DHS) program. However, the geostatistical analyses ignore the fact that GPS coordinates in DHS surveys are jittered for privacy purposes. 
We demonstrate the need to account for this jittering, and we propose a computationally efficient approach that can be routinely applied. We use the new method to analyse the prevalence of completion of secondary education for 20--49 year old women in Nigeria in 2018 based on the 2018 DHS survey. The analysis demonstrates substantial changes in the estimates of spatial range and fixed effects compared to when we ignore jittering. Through a simulation study that mimics the dataset, we demonstrate that accounting for jittering reduces attenuation in the estimated coefficients for covariates and improves predictions. The results also show that the common approach of averaging covariate values in windows around the observed locations does not lead to the same improvements as accounting for jittering.
\end{abstract}

\keywords{
Jittering, \and DHS surveys, \and Demographic and health indicators, \and Geostatistical analysis, \and Template Model Builder (TMB).}

\section{Introduction \label{sec:introduction}}

Fine-scale spatial estimation of demographic and health indicators has become commonplace \citep{burstein2019mapping,utazi2019mapping,local2021mapping}. This paper is focused on prevalences, which include many important indicators such as completion of secondary education, neonatal mortality, and vaccination coverage \citep{fuglstad2021two}. For low- and middle-income countries (LMICs), the household surveys conducted by the 
Demographic and Health Surveys (DHS) Program are a crucial data source. Geographic information in DHS data is given through GPS coordinates, which describe centres of clusters of households. However, cluster centres are randomly displaced by up to $10\,\mathrm{km}$ before being published in order to protect participants' privacy \citep{DHSspatial07}. We refer to such small random displacements as \emph{jittering} of the GPS coordinates.

In global health, it is common practice to ignore jittering and estimate risk using a standard geostatistical model with a binomial likelihood. The latent spatial variation in risk is modelled as the combination of raster- and distance-based covariates and a Gaussian random field (GRF).
However, covariate values extracted from rasters can vary widely on the distance scale of jittering. Using the covariate value at the jittered location instead of the original location induces a non-standard form of measurement error \citep{gustafson-2003}. This may in turn lead to attenuation of effect estimates and errors in uncertainty. Furthermore, not accounting for the positional uncertainty for the GRF, artificially reduces estimated spatial dependency and may reduce predictive power as well \citep{cressie2003spatial,fanshawe2011spatial, fronterre2018geostatistical}.

To address uncertainty in covariates,
\citet{perez2013guidelines, perez2016influence} suggested 1) to use regression calibration in the context of distance-based covariates \citep{warren2016influenceOne}, and 2) to average spatial covariates within a 5 km buffer zone for continuous and categorical rasters. However, this approach does not address the issue of attenuation of associations. \citet{fanshawe2011spatial} proposed a Bayesian approach to account for positional uncertainty for the GRF, but did not propagate uncertainty in the covariates, and only used Gaussian likelihoods that are not applicable to prevalences. The
approach was also computationally expensive, but \citet{fronterre2018geostatistical} made the approach computationally efficient and demonstrated its applicability
to analyse malnutrition based on DHS data.

Recently, \citet{wilson2021estimation} formulated a full geostatistical model for DHS data that includes an observation model for the jittered GPS coordinates, and estimated the model with integrated nested Laplace approximations (INLA) \citep{rue2009approximate} within Markov chain Monte Carlo (MCMC) \citep{gomez2018markov}. Their approach addresses the effect of positional uncertainty on both the spatial covariates and the GRF, but was computationally expensive with 1000 MCMC iterations requiring 52 hours in their simulation study. 
\citet{altay2022accounting} proposed a similar model as \citet{wilson2021estimation}, but used a more efficient inference scheme with computation time being measured in minutes instead of hours. Their approach was made possible through an approximation of the likelihood, the stochastic partial differential equations (SPDE) approach \citep{Lindgren:etal:11}, and Laplace approximations through template model builder (TMB) \citep{JSSv070i05}.

The simulation study in \citet{altay2022accounting} revealed that small spatial ranges for the GRF or larger jittering than the DHS scheme were required to see substantial improvements with the new approach over ignoring jittering.
\citet{altay2022accounting} focused on the impact of jittering on the GRF, and lacked raster- and distance-based covariates. 
Such covariates are far more variable at small spatial scales than a smoothly varying GRF. The aim of this paper is to extend the approach in \citet{altay2022accounting} to a full generalized geostatistical model for prevalence, and to demonstrate that ignoring 
jittering can lead to attenuation of associations and reduced predictive power when analysing DHS data. We show this via a spatial analysis of the prevalence of secondary education completion among women aged 20--49 in 2018 based on the 2018 Nigeria DHS (NDHS2018) \citep{NDHS2018}.

In addition to the new approach, which adjusts for jittering, and the standard approach, which ignores jittering, we consider the common approach of averaging covariates in $5\,\text{km}\times 5\,\text{km}$ windows around the provided GPS coordinates \citep{perez2013guidelines, perez2016influence}. The three methods cannot be compared with cross-validation since the true coordinates of the clusters are not known. Therefore, we construct a simulation study that mimics the NDHS2018 dataset to compare them in terms of their ability to estimate parameters and to predict risk at unobserved locations. We use bias and root mean square error (RMSE) to assess parameter estimation, and RMSE and continuous rank probability score (CRPS) \citep{gneiting2007strictly} to assess predictive ability.

We introduce the datasets and variables of interest in Section \ref{sec:dataDescription}. Then, in Section \ref{sec:methods}, we describe the new approach that adjusts for jittering, and discuss its implementation. In Section \ref{sec:simulation}, we evaluate parameter estimation and prediction with the different methods through a simulation study that mimics the prevalence of secondary education completion. 
Then, in Section \ref{sec:application}, we demonstrate the differences between adjusting and not adjusting for jittering when analysing the prevalence of secondary education completion for women aged 20--49 in Nigeria. The paper ends with discussion and conclusions in Section \ref{sec:discussion}. The code used in the paper can be found in the GitHub repository \url{https://github.com/umut-altay/GeoAdjust}.

\section{Data sources and variables of interest\label{sec:dataDescription}}

Our outcome of interest is completion of secondary education, which is as an indicator of social well-being and life outcome \citep{lewin2008}. Rates vary strongly between women and men, but also between urban and rural areas. According to \citet{unesco2019}, only 1\% of the poorest girls in low income countries will complete secondary education. If a girl completes secondary education, the risk of HIV infection is reduced by about 50\% \citep{unaids2019}.

We consider the prevalence of secondary education completion for women aged 20--49 years in Nigeria in 2018. The lower bound was chosen because younger women may not have completed secondary education yet, and the upper limit was chosen since older women are not available in the DHS surveys. The year 2018 was chosen since this corresponds to the most recent DHS survey in Nigeria, NDHS2018.

Nigeria is an LMIC with a population of more than 200 million, and NDHS2018 has data collected from 1389 clusters with responses from 33,398 women aged 20--49, where 15,621 of them reported that they had completed their secondary education. In this paper, we use the 1380 clusters with valid GPS coordinates, which have responses from 33,193 women aged 20--49 where 15,490 reported that they had completed their secondary education.

\begin{figure}[H]
    \centering
    \subfigure[ Admin2 areas and cluster locations.\label{fig:geography:admin2}]{\includegraphics[height=4.55cm]{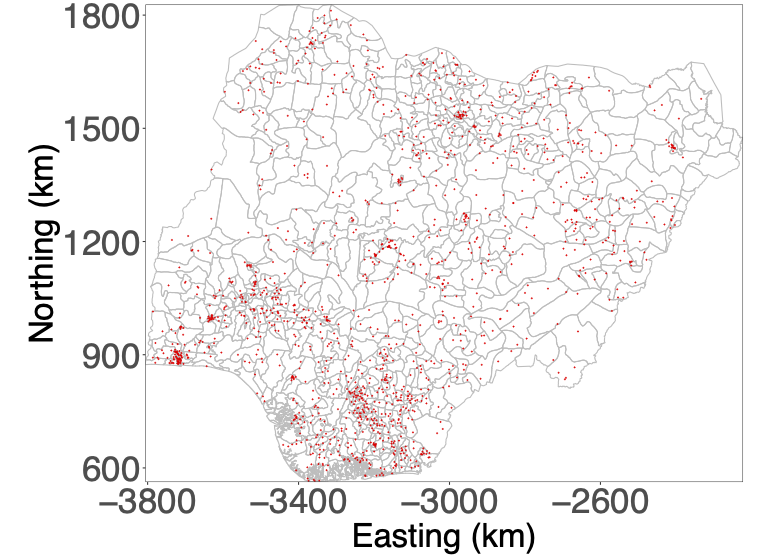}
    }
    \subfigure[Direct estimates for admin1 \label{fig:geography:admin1}]{\includegraphics[height=5cm]{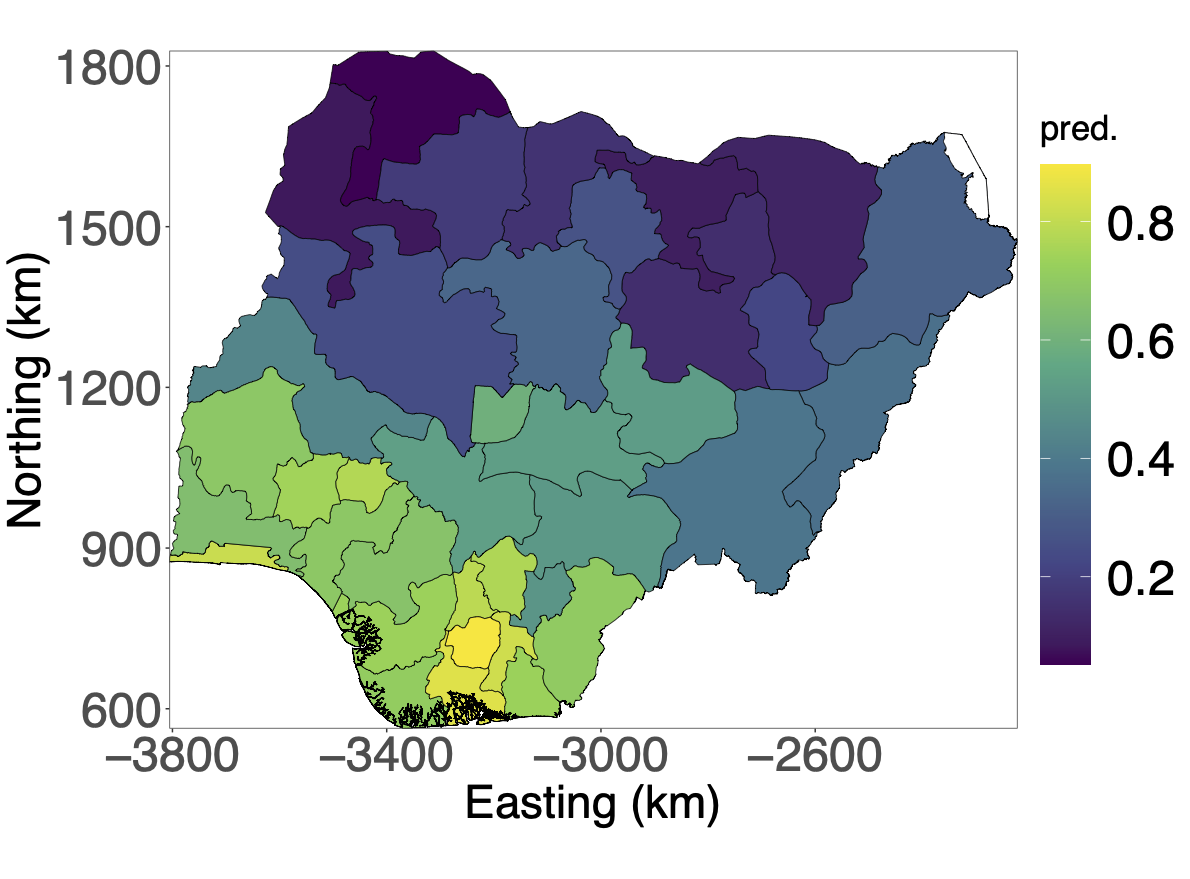}} 
    \caption{Maps of a) admin2 areas and cluster locations for NDHS2018, and b) admin1 areas with direct estimates.}
    \label{fig:geography}
\end{figure}

Figure \ref{fig:geography:admin1} shows the  direct estimates \citep{rao:molina:15} computed based on the data and the survey design for the 36 states and one federal capital territory. The corresponding uncertainty is expressed through the coefficient of variation (CV). These 37 areas are the first administrative level (admin1), and the results show considerable variation between areas. The CVs increase when direct estimates are calculated at finer spatial scales due to observations needing to be distributed amongst more areas, resulting in smaller sample sizes per area. The second administrative level (admin2) for Nigeria, for example, consists of the 774 local government areas shown in Figure \ref{fig:geography:admin2}. The red dots in the figure indicate the 1380 clusters with GPS coordinates within Nigeria available in NHDS2018. The national boundary, admin1 boundaries and admin2 boundaries are based on GADM version 4.0 \citep{GADM}.

We expect the prevalence of completion of secondary education to be closely related to
the access to educational resources, such as technological infrastructure, schools and teachers. We consider five
spatial covariates: population count (PopD) \citep{pop},
travel time to nearest city (CityA) \citep{weiss2018global}, elevation (Elev) \citep{elev}, distance to nearest river or lake (DistW) \citep{riverLake}, and urbanicity ratio \citep{pesaresi2016operating}. 
For UrbR, we scale the original covariate to be on a zero to one scale, and for the other four covaraties, we use a $\log(1+x)$-transformation and then center and standardize the covariate rasters accross the pixels.
The information about the covariate rasters and figures is summarized in Table \ref{tab:covTable}. The covariate rasters are available at different resolutions, and have not been resampled or aggregated to the same resolution since this would involve an extra preprocessing step that would induce misalignment error and potentially add ecological bias \citep{greenland:morgenstern:89}.

\renewcommand{\arraystretch}{0.8}
\begin{table}
    \centering
    \caption{Summary of covariate rasters providing name, description and figure. CityA, Elev, DistW and UrbR are transformed, while UrbR is not.\label{tab:covTable}}
    \vspace{.1cm}
    \begin{tabular}{lll}
    \textbf{Name} & \textbf{Description} & \textbf{Figure} \\
    \hline
    PopD & Population count  ($250\, \mathrm{m} \times 250\, \mathrm{m}$)  & \ref{fig:pop} \\
    CityA & Travel time in minutes ($1\, \mathrm{km} \times 1\, \mathrm{km}$) & 
\ref{fig:travelTime} \\
    Elev & Elevation in meters ($1\, \mathrm{km} \times 1\, \mathrm{km}$)  &\ref{fig:elev}\\
    DistW & Distance to nearest river or lake in degrees ($1\, \mathrm{km} \times 1\, \mathrm{km}$) & 
 \ref{fig:dist}\\
 UrbR & Urbanicity ratio ($250\, \mathrm{m} \times 250\, \mathrm{m}$) & \ref{fig:urb}
    \end{tabular}
\end{table}
\renewcommand{\arraystretch}{1.0}

\begin{figure}[H]
    \centering
    \vspace{-.3cm}
    \subfigure[ \label{fig:pop}]{\includegraphics[width=2.75in]{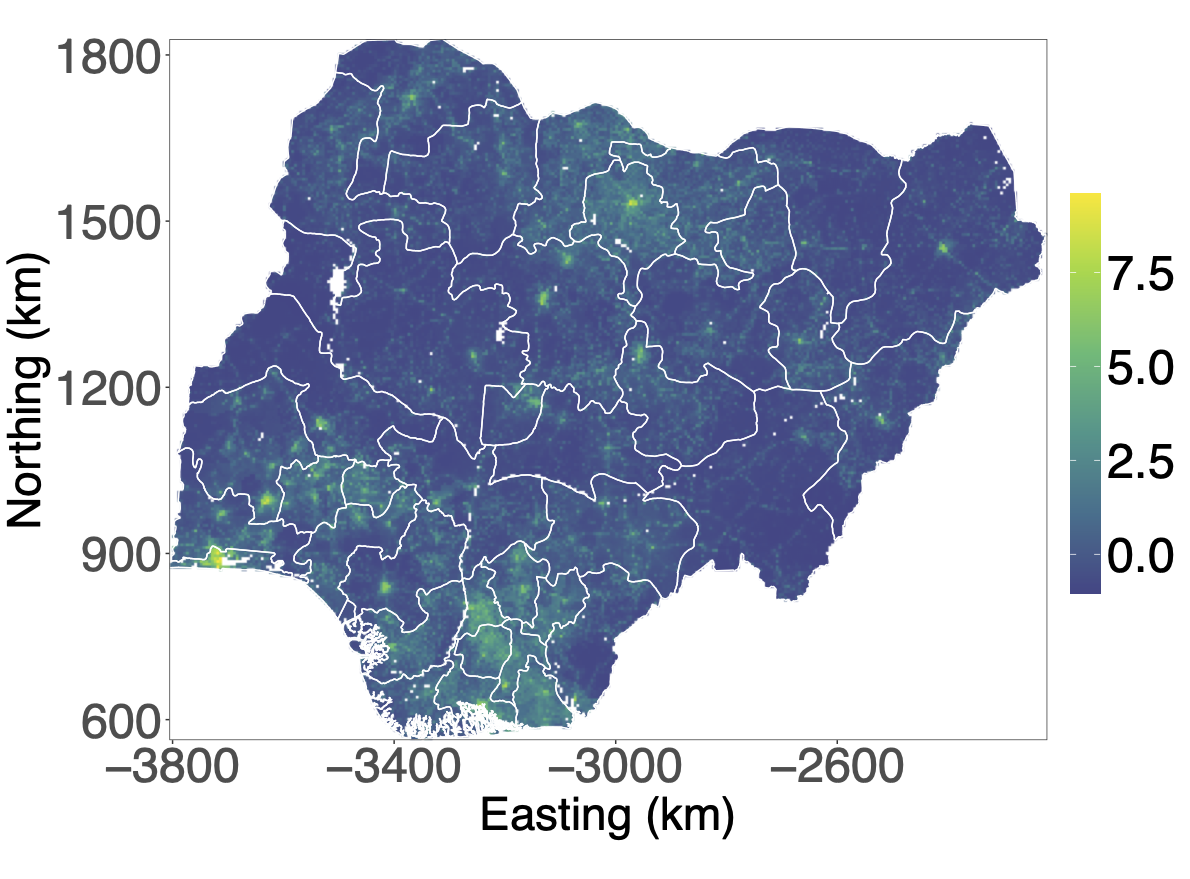}}
    \subfigure[ \label{fig:travelTime}]{\includegraphics[width=2.75in]{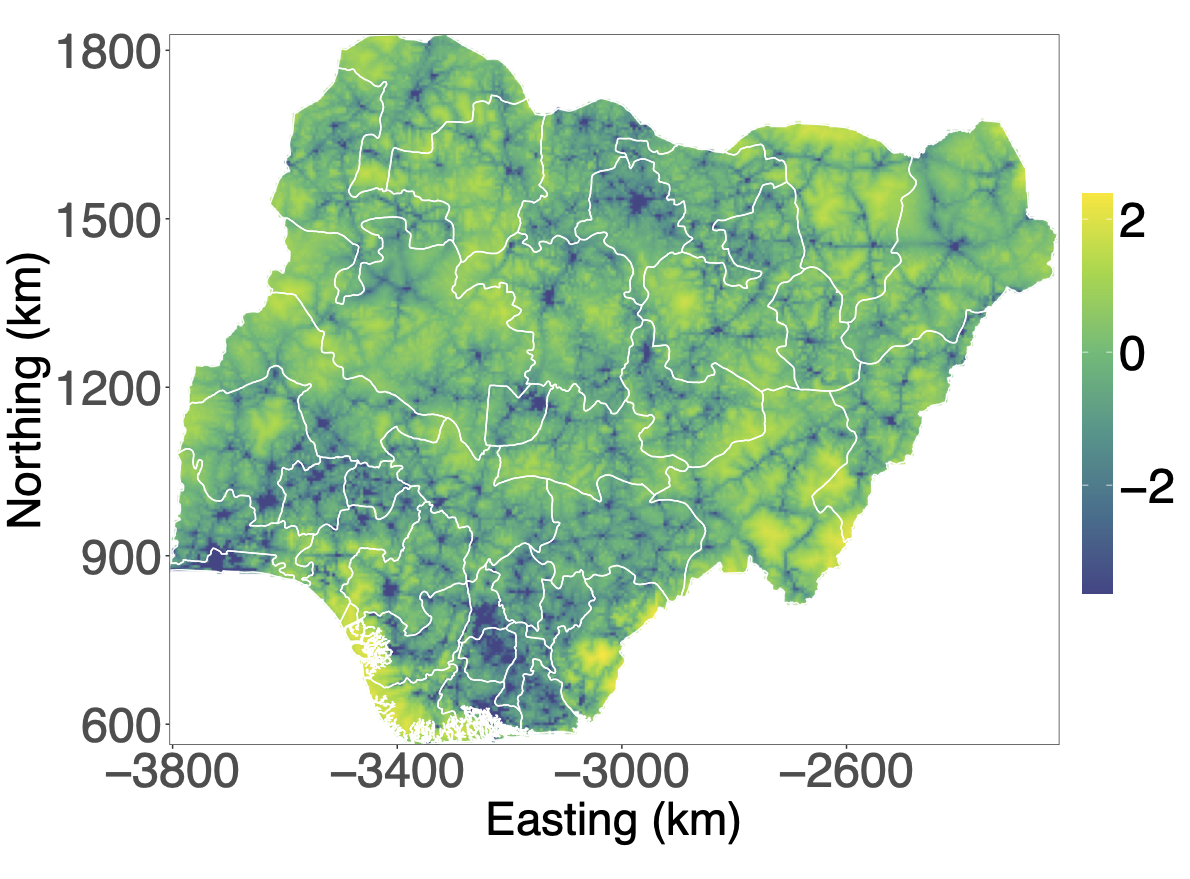}}
    \medskip
    \subfigure[ \label{fig:elev}]{\includegraphics[width=2.75in]{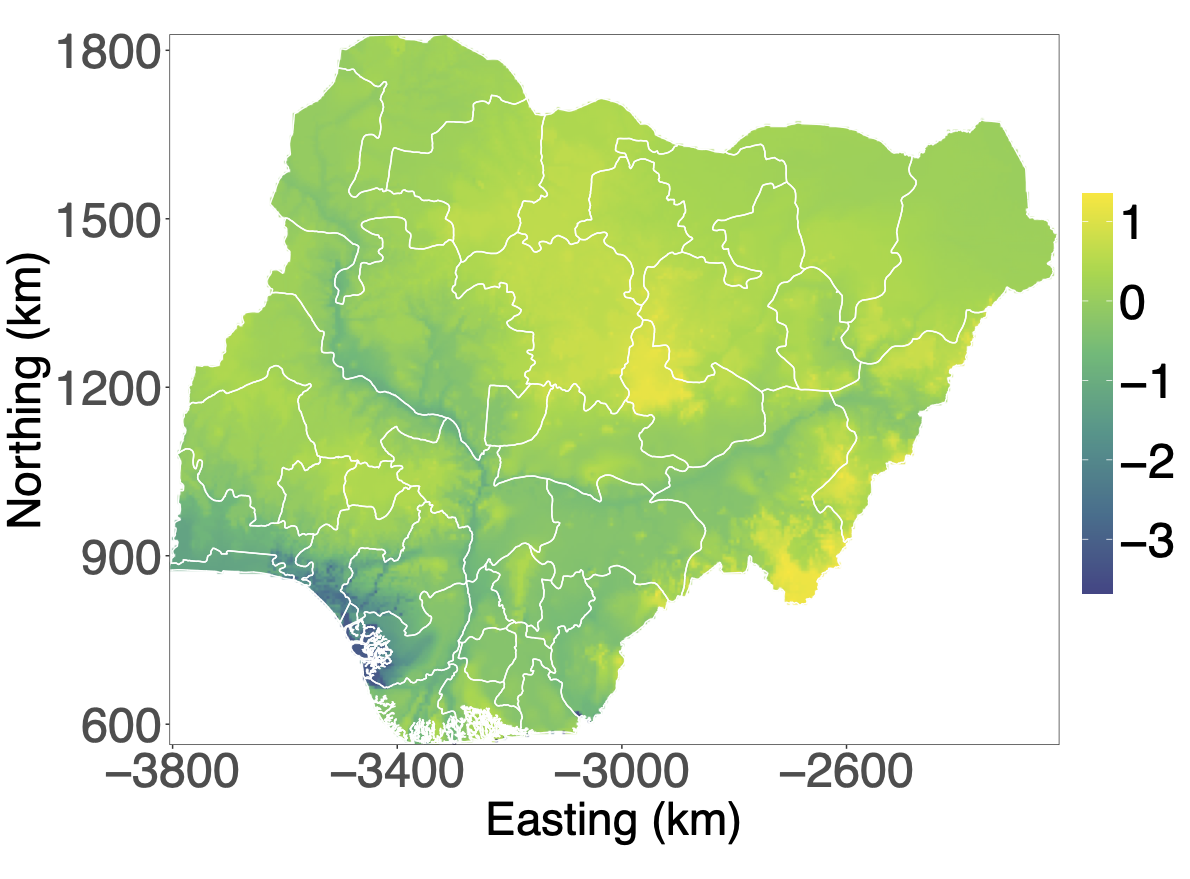}} 
    \medskip
    \subfigure[ \label{fig:dist}]{\includegraphics[width=2.75in]{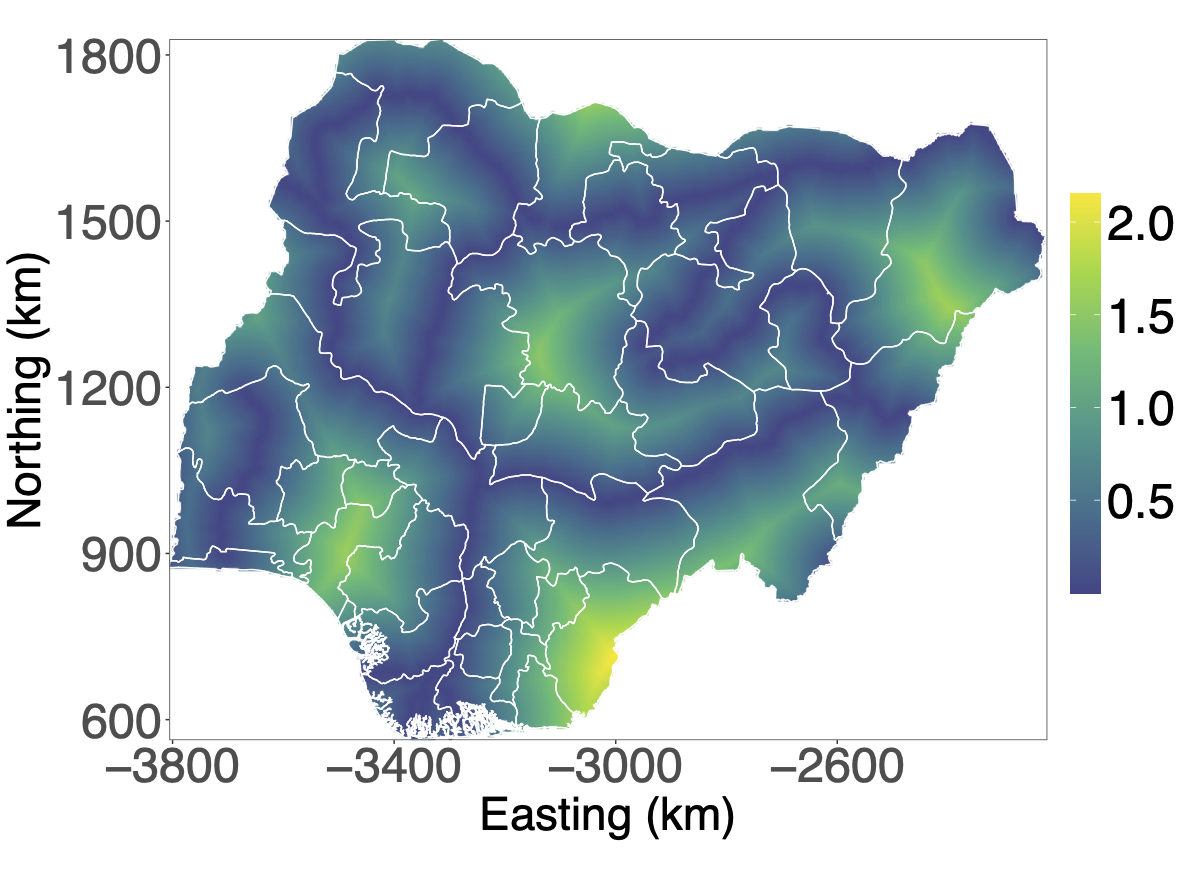}} 
    \subfigure[ \label{fig:urb}]{\includegraphics[width=2.75in]{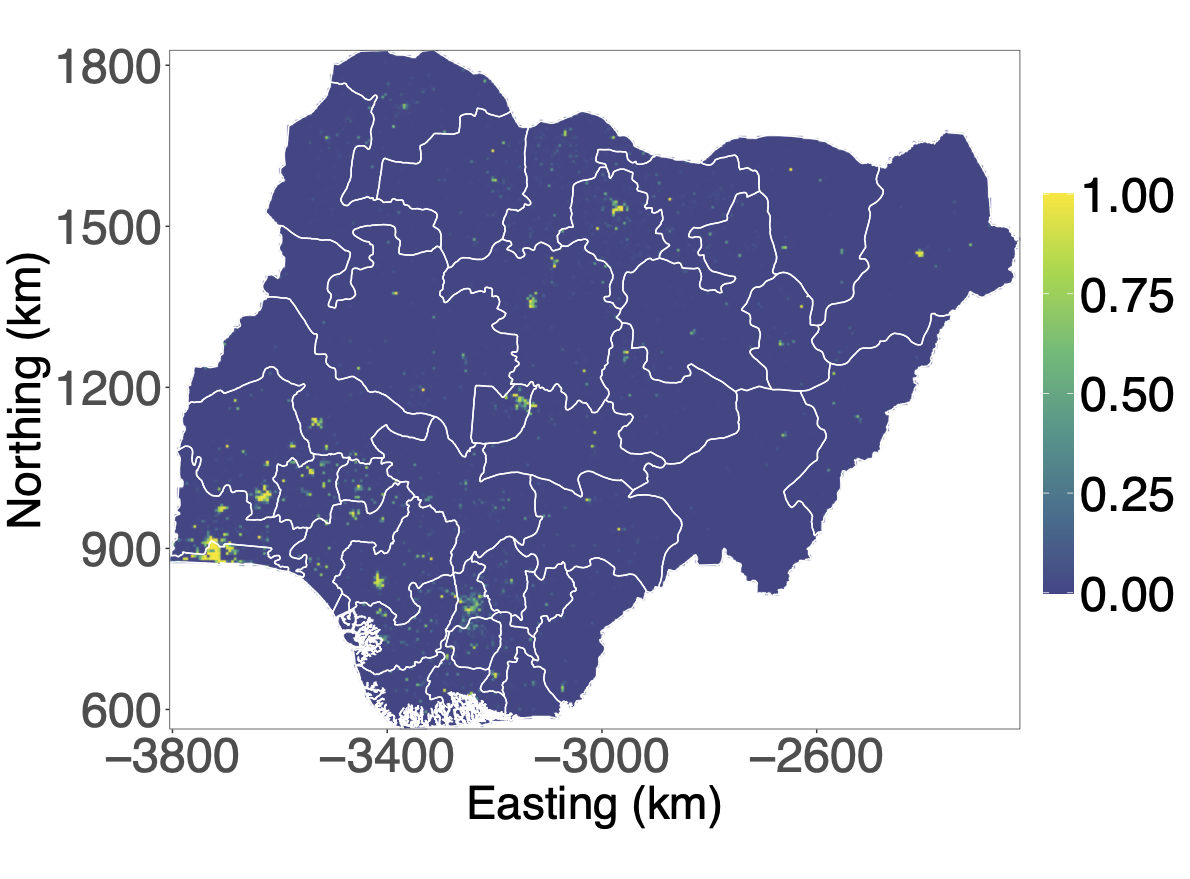}}

    \caption{Transformed covariate rasters for Nigeria: a) population density, b) travel time to the nearest city center, c) elevation, d) minimum distance to nearest river or lake, and e) urbanicity ratio.}
    \label{fig:fiveRasters}
\end{figure}

We aim to map spatial variation at a continuous spatial scale through a geostatistical model. A key challenge with using the above covariates is that the locations shown in Figure \ref{fig:geography:admin2} are not the true locations. The true locations have been randomly displaced while respecting the admin2 borders. This means that we can only imprecisely extract covariates from the rasters shown in Figure \ref{fig:fiveRasters}. The new method to account for the uncertainty in locations is described in Section \ref{sec:methods}.

\section{Adjusting for jittering in a geostatistical model \label{sec:methods}}

\subsection{Notation for DHS data}
For a given country and DHS household survey, $C$ clusters are visited. These clusters constitute small geographic areas and are collections of households. A total of $n_c$ people at risk are observed and $y_c\leq n_c$ individuals have positive outcomes for clusters $c = 1, 2, \ldots, C$, . The reported GPS coordinates of the cluster centres are $\boldsymbol{s}_c\in\mathbb{R}^2$, $c = 1, \ldots, C$. These locations are not the true GPS coordinates, but the jittered GPS coordinates. Additionally, the urban/rural designation is known for each visited cluster.

\subsection{Geostatistical model\label{sec:model}}
\subsubsection{Model for spatial variation in risk}
We envision a spatially varying risk, $r(\cdot)$, for the country of interest $\mathcal{D}\subset \mathbb{R}^2$ modelled through
\[
    r(\boldsymbol{s}) = \mathrm{logit}^{-1}(\eta(\boldsymbol{s})) = \mathrm{logit}^{-1}(\boldsymbol{x}(\boldsymbol{s})^\mathrm{T}\boldsymbol{\beta}+u(\boldsymbol{s})), \quad \boldsymbol{s}\in\mathcal{D},
\]
where $\boldsymbol{x}(\cdot)$ is a $p$-dimensional vector of covariates, $\boldsymbol{\beta}$ is a $p$-dimensional vector of covariate effect sizes, and $u(\cdot)$ is a Matérn GRF. The Matérn covariance function with smoothness $\nu=1$ is parametrized as
\[
C_\mathrm{M}(\boldsymbol{s}_1, \boldsymbol{s}_2; \sigma_\mathrm{S}^2, \rho_\mathrm{S})=\sigma_\mathrm{S}^2\left(\sqrt{8}\frac{||\boldsymbol{s}_1-\boldsymbol{s}_2||}{\rho_\mathrm{s}}\right) \mathrm{K}_1\left(\sqrt{8}\frac{||\boldsymbol{s}_1-\boldsymbol{s}_2||}{\rho_\mathrm{s}}\right), \quad \boldsymbol{s}_1, \boldsymbol{s}_2 \in \mathcal{D},
\]
where $\sigma_\mathrm{S}^2$ is the marginal variance, $\rho_\mathrm{S}$ is the spatial range, and $\mathrm{K}_1$ is the modified Bessel function of the second kind, order 1.

\subsubsection{Unadjusted model}
\label{sec:unadjusted}
When jittering is ignored, the reported cluster locations $\boldsymbol{s}_1, \ldots, \boldsymbol{s}_C$ are treated as the true locations. This gives the unadjusted observation model:
\begin{align}
\label{eqn:unadjObsModel}
\begin{split}
y_c | r_c,n_c &\sim \text{Binomial}(n_c, r_c),\\
    r_c &= r(\boldsymbol{s}_c) = \mathrm{logit}^{-1}( \eta(\boldsymbol{s}_c)),
\end{split}
\end{align}
where $r_c$ is the risk in cluster $c$, for $c = 1, \ldots, C$. 

\subsubsection{Adjusted model}
Let $\boldsymbol{s}_1^*, \ldots, \boldsymbol{s}_C^* \in\mathcal{D}$ denote the true locations corresponding to the jittered locations $\boldsymbol{s}_1, \ldots, \boldsymbol{s}_C$. The adjusted observation model is, 
\begin{align}
\label{eqn:adjObsModel}
\begin{split}
y_c | r_c,n_c &\sim \text{Binomial}(n_c, r_c), \quad \boldsymbol{s}_c|\boldsymbol{s}_c^*\sim \pi_{\mathrm{Urb}[c]}(\boldsymbol{s}_c|\boldsymbol{s}_c^*),\\
    r_c | \boldsymbol{s}_c^* &= r(\boldsymbol{s}_c^*) = \mathrm{logit}^{-1}( \eta(\boldsymbol{s}_c^*)),
\end{split}
\end{align}
where $r_c$ is the risk in cluster $c$, and $\mathrm{Urb}[c]\in\{\mathrm{U}, \mathrm{R}\}$ corresponds to the cluster's urban (U) or rural (R) designation, for $c = 1, \ldots, C$. In this observation model, both $y_c$ and $\boldsymbol{s}_c$ are treated as observed quantities. The unobserved true locations $s_c^*$ are treated as random quantities and assigned a uniform prior $s_c^*\sim \mathcal{U}(A(\boldsymbol{s}_c))$, where $A(\boldsymbol{s})\in\{1, \ldots, K\}$ denotes the administrative region containing location $\boldsymbol{s}\in\mathcal{D}$ that a cluster location must be jittered within and $K$ is the number of such administrative regions. For example, the NDHS2018 jittered cluster locations must be in the same admin2 area as the associated true cluster locations. This implies that we treat all true cluster locations $\boldsymbol{s}_c^*$ in the corresponding admin2 area and within the maximum jittering distance of $\boldsymbol{s}_c^*$ as equally likely \emph{a priori}.

The jittering distributions $\pi_\mathrm{U}$ and $\pi_\mathrm{R}$ follow from the (known) DHS jittering scheme. Then for an urban cluster $c$, which can be jittered up to $2\, \mathrm{km}$, the jittering distribution is 
\[
    \pi_\mathrm{U}(\boldsymbol{s}_c|\boldsymbol{s}_c^*) \propto \frac{\mathbb{I}(A(\boldsymbol{s}_c) =A( \boldsymbol{s}_c^*))\cdot\mathbb{I}( d(\boldsymbol{s}_c,\boldsymbol{s}_c^*)<2)}{ d(\boldsymbol{s}_c,\boldsymbol{s}_c^*)}, \quad \boldsymbol{s}_c \in \mathcal{D},
\]
where $d(\boldsymbol{s}_c,\boldsymbol{s}_c^*)$ is the distance in kilometers between $\boldsymbol{s}_c$ and $\boldsymbol{s}_c^*$, and $\mathbb{I}$ is the indicator function. For a rural cluster $c$, which can be jittered up to $5\, \mathrm{km}$ except for $1\%$ of clusters jittered up to $10\, \mathrm{km}$, the jittering distribution is:
\[
    \pi_\mathrm{R}(\boldsymbol{s}_c|\boldsymbol{s}_c^*) \propto \frac{\mathbb{I}(A(\boldsymbol{s}_c) =A( \boldsymbol{s}_c^*))}{d(\boldsymbol{s}_c, \boldsymbol{s}_c^*)} \left[\frac{99\mathbb{I}(d(\boldsymbol{s}_c,\boldsymbol{s}_c^*)<5)}{100} + \frac{\mathbb{I}( d(\boldsymbol{s}_c,\boldsymbol{s}_c^*)<10)}{100}\right], \quad \boldsymbol{s}_c \in \mathcal{D}.
\]

\subsubsection{Priors}
We assume linear covariate associations, and use the prior $\boldsymbol{\beta}\sim\mathcal{N}_p(\boldsymbol{0}, 25\mathbf{I}_p)$. The range, $\rho_\mathrm{S}$, and marginal variance, $\sigma_\mathrm{S}^2$, of the Matérn GRF is assigned a penalised complexity (PC) prior \citep{fuglstad:etal:19a}. This requires selecting two hyperparameters: the \emph{a priori} median of range $R_0$, and the \emph{a priori} 95th percentile of marginal standard deviation $S_0$.

\subsection{Implementation \label{sec:Implementation}}
\subsubsection{Inference scheme\label{sec:Implementation:Inf}}
The observation model in Equation \eqref{eqn:adjObsModel} can be written as,
\begin{align}
    \pi(y_c, \boldsymbol{s}_c|\eta(\cdot)) &= \int_{\mathbb{R}^2} \pi(y_c, \boldsymbol{s}_c| \eta(\cdot), \boldsymbol{s}_c^*) \pi(\boldsymbol{s}_c^*) \ \mathrm{d}\boldsymbol{s}_c^* \notag \\
    &= \int_{\mathbb{R}^2} \pi(y_c| \eta(\boldsymbol{s}_c^*)) \pi_{\mathrm{Urb}[c]}(\boldsymbol{s}_c| \boldsymbol{s}_c^*) \pi(\boldsymbol{s}_c^*) \ \mathrm{d}\boldsymbol{s}_c^*, \label{eq:adjObsMod2}
\end{align}
for $c = 1, \ldots, C$.  Let $\boldsymbol{\theta} = (\log(\sigma_\mathrm{S}^2), \log(\rho_\mathrm{S}))$. We propose an empirical Bayes
approach:
\begin{itemize}
    \item \textbf{Step 1:} Calculate the maximum a posteriori (MAP) estimate, $\hat{\boldsymbol{\theta}}$, of $\boldsymbol{\theta}$ using  $\pi(\boldsymbol{\theta}|y_1, \ldots, y_C, \boldsymbol{s}_1, \ldots, \boldsymbol{s}_C)$.
    \item \textbf{Step 2:} Extract inference about $\boldsymbol{\beta}$ from $\pi(\boldsymbol{\beta}|y_1, \ldots, y_C, \boldsymbol{s}_1, \ldots, \boldsymbol{s}_C, \boldsymbol{\theta}=\hat{\boldsymbol{\theta}})$.
    \item \textbf{Step 3:} Estimate risk $r(\boldsymbol{s})$ at location $\boldsymbol{s}$ using \[   \pi(r(\boldsymbol{s})|y_1, \ldots, y_C, \boldsymbol{s}_1, \ldots, \boldsymbol{s}_C, \boldsymbol{\theta}=\hat{\boldsymbol{\theta}}).
        \]
\end{itemize}
Two key components are combined for rapid inference: 
the SPDE approach to approximate the Matérn GRF \citep{Lindgren:etal:11}, and Template Model Builder (TMB) for empirical Bayesian inference \citep{JSSv070i05}.

\subsubsection{SPDE approach}
For each cluster $c$, the true location $\boldsymbol{s}_c^*$ is not known, and the observation model in Equation \eqref{eq:adjObsMod2} involves the spatial field $u(\cdot)$ at all locations that are compatible with the jittered location $\boldsymbol{s}_c$. If we replace the integral in Equation \eqref{eq:adjObsMod2} by a integration scheme using $N_\mathrm{Int}$ integration points, we need to evaluate the spatial field at $C\cdot N_\mathrm{Int}$ locations. A standard implementation of the Matérn model would result in a dense $C\cdot N_\mathrm{Int} \times C\cdot N_\mathrm{Int}$ matrix and make computations infeasible even for a few locations.

The SPDE approach \citep{Lindgren:etal:11} overcomes this issue by approximating the Matérn GRF that results in a sparse precision matrix. First, the area of interest is triangulated with a triangulation consisting of $m$ nodes. Then the GRF $u(\cdot)$ is approximated by
\begin{equation}
\tilde{u}(\boldsymbol{s}) = \sum_{i=1}^m w_i \phi_i(s),\label{eq:SPDE:basis}
\end{equation}
where $\phi_i(\cdot)$ are pyramidal basis functions and  $\boldsymbol{w} = (w_1 \ \ldots \ w_m)^\mathrm{T}$
are weights for the basis functions. The SPDE approach results in a distribution $\boldsymbol{w}\sim\mathcal{N}_m(\boldsymbol{0}, \mathbf{Q}(\boldsymbol{\theta})^{-1})$, where $\mathbf{Q}(\boldsymbol{\theta})$ is sparse.

From Equation \eqref{eq:SPDE:basis}, the value at any location is a linear transformation $\tilde{u}(\boldsymbol{s}) = \boldsymbol{a}(\boldsymbol{s})^\mathrm{T}\boldsymbol{w}$, $\boldsymbol{s}\in\mathcal{D}$, where $\boldsymbol{a}(\boldsymbol{s})\in\mathbb{R}^m$ is sparse with at most three nonzero elements depending on the location $\boldsymbol{s}$. This means that the spatial field can be evaluated at a large number of locations quickly.

The SPDE is given by
\[
    (\kappa^2-\Delta)(\tau u(\boldsymbol{s})) = \mathcal{W}(\boldsymbol{s}), \quad \boldsymbol{s}\in \tilde{\mathcal{D}},
\]
where $\kappa>0$ and $\tau>0$ are related to marginal variance and range, $\Delta$ is the Laplacian, $\mathcal{W}(\cdot)$ is standard Gaussian white noise, and $\tilde{\mathcal{D}}\supset\mathcal{D}$ is an extended domain to reduce boundary effects. We use Neumann boundary conditions to make the problem well defined, and following \citet{Lindgren:etal:11}, the effective range and marginal variance are calculated from the SPDE parameters as
\[
    \rho_\mathrm{S} = \frac{\sqrt{8}}{\kappa}\quad \text{and}\quad \sigma_\mathrm{S}^2 =\frac{1}{4\pi\tau^2\kappa^2}.
\]

\subsubsection{Template Model Builder}
\label{sec:method:TMB}
We implement the empirical Bayesian inference scheme by employing the built-in auto-differentiation and Laplace approximations of the TMB R package. Unlike sampling based MCMC methods, TMB uses numerical integration, through Laplace approximations, to perform inference. The auto-differentiation is used to speed up
the Laplace appoximations. See \citet{JSSv070i05} for the details.

In TMB, one can compute arbitrary likelihoods, and we can approximate the likelihood in Equation \eqref{eq:adjObsMod2} through the integration scheme
\begin{equation}
    \pi(y_c, \boldsymbol{s}_c| \eta(\cdot)) \propto
    \sum_{k = 1}^K
    \alpha_{k}\pi(y_c| \eta(\boldsymbol{s}_{c,k}^*))\pi_{\mathrm{Urb}[c]}(\boldsymbol{s}_{c}|\boldsymbol{s}_{c,k}^*)\pi(\boldsymbol{s}_{c,k}^*),
    \label{eq:intScheme}
\end{equation}
where $\alpha_1, \ldots, \alpha_K$ are integration weights. More details are available in \citet{altay2022accounting}.  Critically, the integration scheme in Equation \eqref{eq:intScheme} involves 
\[
\eta(\boldsymbol{s}_{c,k}^*) = \boldsymbol{x}(\boldsymbol{s}_{c,k}^*)^\mathrm{T}\boldsymbol{\beta} + u(\boldsymbol{s}_{c,k}^*), \quad  k = 1, \ldots, K, \quad c = 1, \ldots, C.
\]
Based on the known jittering distribution, we construct the integration scheme with rings of integration points around each cluster center. The observed cluster center is the first integration point, and we use 5 and 10 rings for the clusters that are located within urban and rural administrative areas, respectively. Each ring consists of 15 angularly equidistant integration points.

Throughout this paper we consider the three approaches shown in Table \ref{tab:methods}. UnAdj denotes the traditional model that does not adjust for jittering, Smoothed denotes a model where the covariates have been averaged over $5\, \text{km}\times 5\, \text{km}$ windows around each location \citep{perez2013guidelines,perez2016influence} and FullAdj denotes the new geostatistical model that fully adjust for the positional uncertainty.

\begin{table}
    \centering
    \caption{The three approaches considered in the paper.\label{tab:methods}}
    \begin{tabular}{ll}
        \textbf{Approach} & \textbf{Description} \\
        \hline
        UnAdj & Traditional geostatistical model. \\
        Smoothed & UnAdj with covariates averaged over $5\, \text{km}\times 5\, \text{km}$ windows. \\
        FullAdj & Geostatistical model adjusting for jittering. \\
    \end{tabular}
\end{table}

\section{Simulation study \label{sec:simulation}}
In this section, we evaluate the three models in Table \ref{tab:methods} using a known data generating model, where the parameters correspond to FullAdj estimated based on completion of secondary education in Section \ref{sec:application}, see Table \ref{tab:realData2columns}. The data generating model is chosen to achieve realistic scenarios, and the aim is to evaluate accuracy in parameter estimation, and accuracy in predictive distributions. The key interest is how these accuracies vary across different strengths of the 
signal from the covariates.

We assume that the true spatial risk varies as
\begin{equation}
    r(\boldsymbol{s}) = \mathrm{logit}^{-1}(\boldsymbol{x}(\boldsymbol{s})^\mathrm{T}\boldsymbol{\beta}+u(\boldsymbol{s})), \quad \boldsymbol{s}\in\mathcal{D},\label{eq:simRisk}
\end{equation}
where $\mathcal{D}$ is Nigeria, $\boldsymbol{x}(\cdot)$ is a 6-dimensional spatially varying vector with 1 as the first element denoting the intercept, and the covariates DistW, CityA, Elev, PopD, and UrbR as the five last elements, $\boldsymbol{\beta} = (\mu, \beta_{\mathrm{DistW}}, \beta_{\mathrm{CityA}}, \beta_{\mathrm{Elev}}, \beta_{\mathrm{PopD}}, \beta_{\mathrm{UrbR}})^\mathrm{T}$, and $u(\cdot)$ is a Matérn GRF. Spatial range $\rho_\mathrm{S}$ and the marginal variance $\sigma_\mathrm{S}^2$ of the GRF are set to the values given in Table \ref{tab:realData2columns} for FullAdj. Then we construct three scenarios according to weaker, the same, and stronger association to the covariates compared to estimated coefficients under FullAdj in Table \ref{tab:realData2columns}:
\begin{itemize}
    \item[1.] \textbf{SignalLow}: $\boldsymbol{\beta}$ is set to $0.5$ times the estimated values.
    \item[2.] \textbf{SignalMed}: $\boldsymbol{\beta}$ is set to $1.0$ times the estimated values.
    \item[3.] \textbf{SignalHigh}: $\boldsymbol{\beta}$ is set to $1.5$ times the estimated values.
\end{itemize}

For each of the three scenarios, we generate $n_\mathrm{sim} = 50$ true risk surfaces, $r(\cdot)$, according to Equation \eqref{eq:simRisk}.

For observations, we follow the design of the NDHS2018 survey. We fix the number of clusters to $C =$ 1,380, and for each cluster $c$, and fix 568 urban locations and 812 rural locations. We also fix the number-at-risk, $n_c$, according to NDHS2018 for $c = 1, \ldots, C$. For each of the 150 true risk surfaces, the true locations $\boldsymbol{s}_1^*, \ldots, \boldsymbol{s}_C^*\in\mathcal{D}$ are simulated as a Poisson point process with intensity proportional to population density. This ensures higher likelihood to place observations at locations with more people, and no locations at locations with no people. Then, for each 
cluster $c = 1, \ldots, C$, we simulate response $y_c | r(\boldsymbol{s}_c^*), n_c \sim \mathrm{Binomial}(n_c, r(\boldsymbol{s}_c^*))$ and observed location $\boldsymbol{s}_c|\boldsymbol{s}_c^*$ according to the DHS jittering scheme. 

We fit the models UnAdj, Smoothed and FullAdj described in Section \ref{sec:methods} using an intercept and the five covariates described above. The PC prior on the Matérn GRF is specified by $\mathrm{P}(\sigma_\mathrm{S} > S_0) = 0.05$ and 
$\mathrm{P}(\rho_\mathrm{S} > R_0) = 0.50$, where $S_0 = 1$ is the 95th percentile of the marginal standard deviation, and $R_0 = 160\, \mathrm{km}$ is the median range. Inference is
performed as described in Section \ref{sec:Implementation}.

Parameter estimation is evaluated by computing the RMSE, $\frac{1}{n_{\mathrm{sim}}}\sum_{b = 1}^{n_{\mathrm{sim}}} (\hat{\theta}^{(b)}-\theta)^2$, and the Bias, $\frac{1}{n_{\mathrm{sim}}}\sum_{b = 1}^{n_{\mathrm{sim}}} (\hat{\theta}^{(b)}-\theta)$, where $\hat{\theta}^{(b)}$ is the posterior mean (or MAP in the case of $\rho_\mathrm{S}$ and $\sigma_\mathrm{S}^2$) for dataset $b$ and $\theta$ is the true value of the coefficient. 
Predictions are evaluated on a fixed set of 1,000 randomly selected locations within Nigeria, where we predict $\eta(\boldsymbol{s}) = \mathrm{logit}(r(\boldsymbol{s}))$ with the posterior median.
These predictions are evaluated by the average RMSE and CRPS defined by
$\int_{\mathbb{R}^2}(F(x)-\mathbb{I}(y \leq x))^2\mathrm{d}x$, where $y$ is the true value and $F(\cdot)$ is the predictive distribution.

The simulation study was run on a a computing server that operates on Linux (Ubuntu 20.04). The computing server
has 28 cores  (2×14-core Xeon 2.6 GHz) and provides 256 GB memory limit per user. It took on average 4 minutes to estimate the parameters $\sigma_\mathrm{S}$ and $\rho_\mathrm{S}$ for FullAdj across the 150 datasets, which is the most time-consuming part of the inference described in Section \ref{sec:Implementation:Inf}.

Table \ref{tab:simStudy} shows the bias and RMSE for each of the parameters and covariate coefficients. The results show that UnAdj and Smoothed performs almost the same for $\sigma_\mathrm{S}^2$ and $\rho_\mathrm{S}$, but that Smoothed gives higher bias and RMSE for the coefficients of the covariates. The latter is a consequence of the fact that smoothing the covariates changes the interpretation of the coefficients. E.g., the association with the urbanicity ratio of a $250\, \text{m} \times 250\, \text{m}$ pixel and the urbanicity ratio of a $5\, \text{km}\times 5\, \text{km}$ pixel are two different things. 

\begin{figure}[H]
\centering
\includegraphics[width=\textwidth]{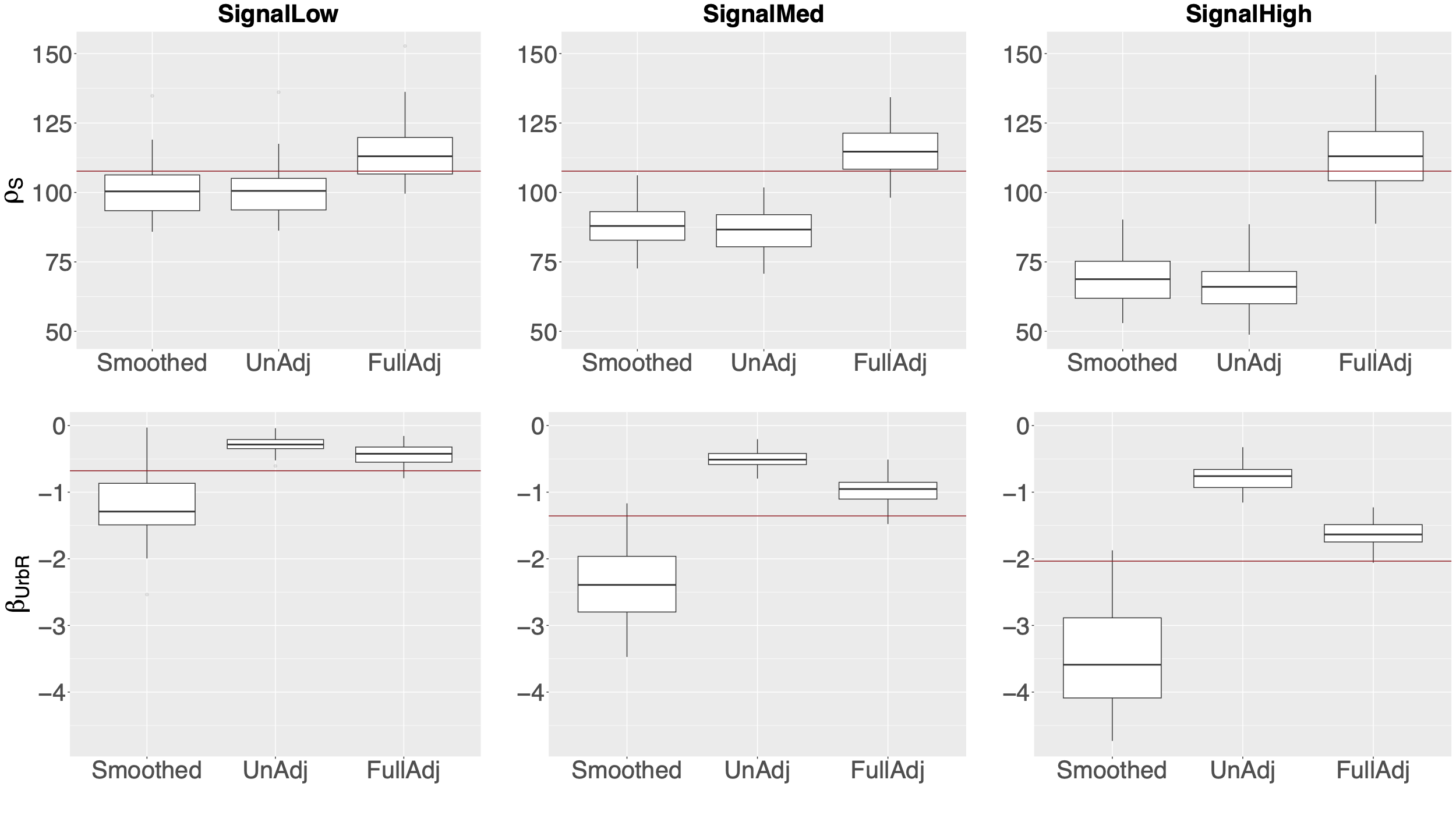}
\caption{Box plots of estimated $\rho_\mathrm{S}$ and $\beta_{\mathrm{UrbR}}$ for SignalLow, SignalMed and SignalHigh. The horizontal red lines show the true parameter value.}
\label{fig:boxplotParameter}
\end{figure}

Figure \ref{fig:boxplotParameter} shows boxplots of estimated $\rho_\mathrm{S}$ and $\beta_\mathrm{UrbR}$ underlining that difference between models is also clear when you consider variation between simulations within the same scenario. There is a clear trend to underestimate the spatial range for UnAdj and Smoothed, and the association with the covariate is too weak for UnAdj. For Smoothed, the model is estimating a coefficient with different interpretation than for the other models due to the averaging of covariate rasters over $5\, \text{km}\times 5\, \text{km}$ windows. The differences are larger for stronger covariate signal levels.

\begin{table}[H]
\centering
    \caption{Bias and RMSE of parameter estimation for SignalMed. \label{tab:simStudy}}
    \vspace{5mm}
\renewcommand{\belowrulesep}{0.05ex}
\begin{adjustbox}{max width=\textwidth}
    \begin{tabular}{cccccccccc}
       \toprule
       & & \multicolumn{8}{c}{\textbf{Parameter}} \\[-5pt]
      & \textbf{Model} & $\rho_\mathrm{S}$ & $\sigma_\mathrm{S}^{2}$ & $\mu$ & $\beta_\mathrm{DistW}$ & $\beta_\mathrm{CityA}$ & $\beta_\mathrm{Elev}$ & $\beta_\mathrm{PopD}$ & $\beta_\mathrm{UrbR}$ \\
      \midrule
\multirow{3}{*}{Bias} & \textbf{UnAdj} & -21.32  & 0.05 & 0.32  & 0.00  & 0.08  & 0.01 & -0.16 & 0.84  \\ 
& \textbf{Smoothed} & -19.97 & 0.05 & 0.21 & 0.14 & -0.22 & 0.04 & 0.15 & -1.03 \\
& \textbf{FullAdj} & 7.51 & 0.009 & 0.27 & 0.01 & 0.02 & 0.02 & -0.09 & 0.37 \\
\midrule

\multirow{2}{*}{RMSE} & \textbf{UnAdj} & 22.48 & 0.09 & 0.42 & 0.27 & 0.09 & 0.19 & 0.16 & 0.85 \\ 
& \textbf{Smoothed} & 21.37 & 0.09 & 0.34 & 0.37 & 0.24 & 0.29 & 0.17 & 1.17 \\
& \textbf{FullAdj} & 11.96 & 0.09 & 0.37 & 0.26 & 0.04 & 0.17 & 0.09 & 0.42 \\
\bottomrule
\end{tabular}
\end{adjustbox}
\renewcommand{\belowrulesep}{0.65ex}
\end{table}

\begin{figure}[H]
\centering
\includegraphics[width=\textwidth]{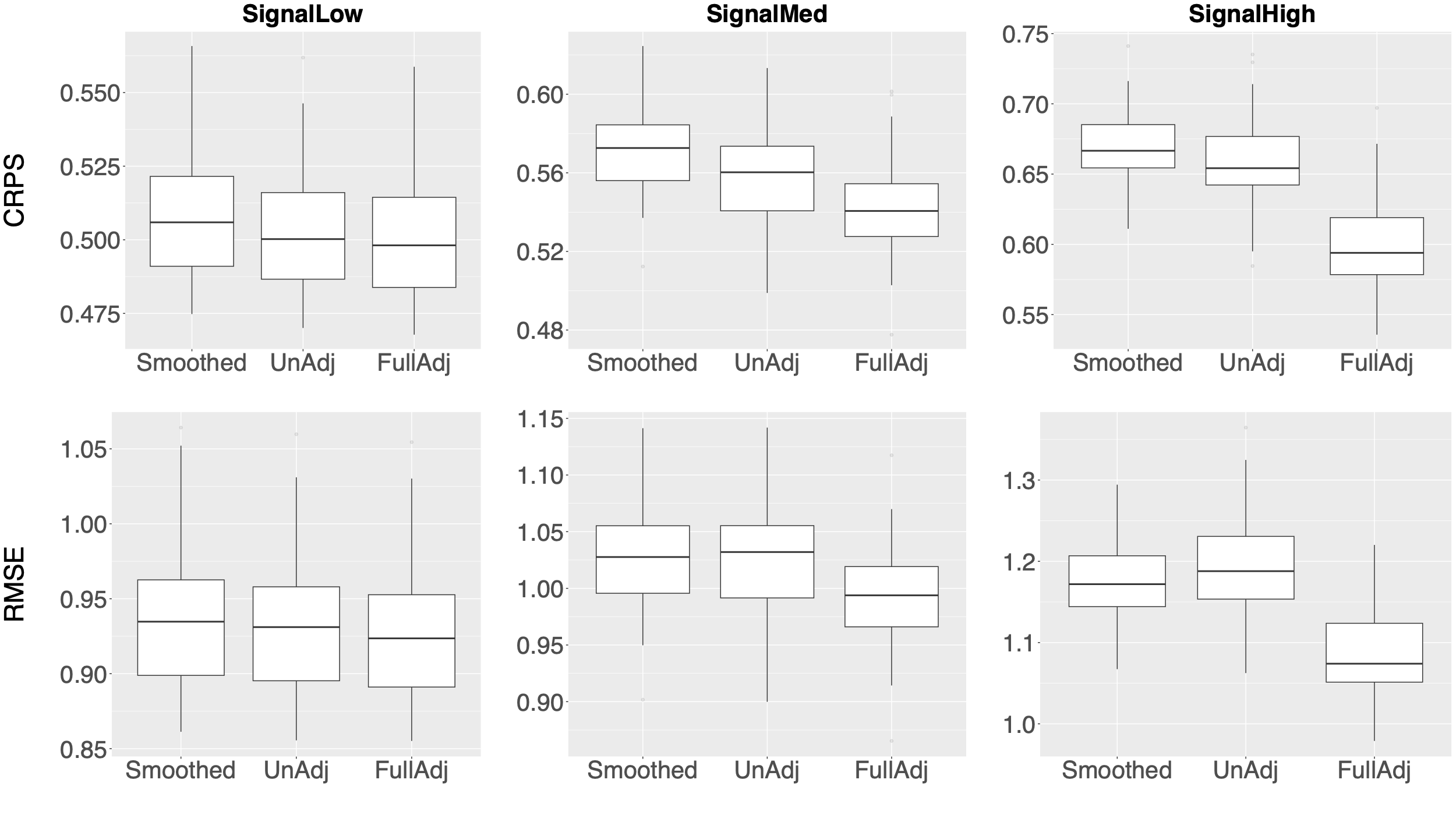}
\caption{Box plots of CRPS and RMSE for predictions for SignalLow, SignalMed and SignalHigh.}
    \label{fig:boxplotPrediction}
\end{figure}

Figure \ref{fig:boxplotPrediction} shows the variation in RMSE and CRPS across datasets for predictions. 
FullAdj and UnAdj perform almost the same in both predictive metrics for SignalLow, but FullAdj is slightly better for SignalMed, and substantially better for SignalHigh. This indicates that the stronger the signal of the spatial covariates, the larger the gain from adjusting for jittering. The figure also indicates that Smoothed does not have better predictive ability than UnAdj.

In addition to the simulation study described above, we performed a simulation study with the true locations fixed to the observed locations of NDHS2018. The goal was to examine the predictive metrics for the specific spatial design of NDHS2018. There were only minor differences between this simulation study, and the one described above, and we, therefore, show these results in Section 2 of the Supplementary materials.  This suggests that we should expect the unadjusted model to underestimate range, and lose predictive accuracy due to reduced association with the covariates.

\section{Analysis of completion of secondary education \label{sec:application}}

In this section, we analyse the spatial variation in prevalence of completion of secondary education among 20--49 year old women in Nigeria in 2018
based on the NDHS2018 using the data sources described in Section 2. Our analysis  
has two aims.
First, to map the fine-scale spatial variation in
completion of secondary education for $5\, \mathrm{km} \times 5\, \mathrm{km}$ pixels,  and for the admin1 areas. Second, to determine the associations between the spatial variation in risk and a set of explanatory spatial covariates.

The NDHS2018 has $C = $ 1,380 clusters with jittered
GPS coordinates available under the same jittering distribution as in Section \ref{sec:model}. For all clusters, the jittering was restricted to stay within the correct admin2 area. In total, 33,193 women aged 20--49 years were interviewed and 15,490 of these had completed secondary education. We use the notation $n_c$ individuals-at-risk, $y_c$ successes, and jittered GPS coordinate $\boldsymbol{s}_c$ for $c = 1, \ldots, C$. The five covariates of interest are introduced in Section 2.

We fit the models UnAdj, Smoothed and FullAdj described in Section \ref{sec:methods} using an intercept and the five covariates described in Section \ref{sec:dataDescription}. We use the PC prior on the Matérn GRF specified by $\mathrm{P}(\sigma_\mathrm{S} > S_0) = 0.05$ and 
$\mathrm{P}(\rho_\mathrm{S} > R_0) = 0.50$, where $S_0 = 1$ is the 95th percentile of the marginal standard deviation, and $R_0 = 160\, \mathrm{km}$ is the median range. Inference is
performed as described in Section \ref{sec:Implementation}.

This application was run on a MacBook Pro (2.4GHz Quad-Core Intel Core i5 and 16 GB memory). For FullAdj, it took 7.8 minutes to estimate parameters, and 4.2 minutes to compute predictions. 

Table \ref{tab:realData2columns} shows the estimated parameters and their corresponding credible interval lengths (except for $\rho_\mathrm{S}$ and $\sigma_\mathrm{S}^2$, which are fixed to their MAP estimates). The results show that $\rho_\mathrm{S}$ is estimated substantially smaller for UnAdj and Smoothed than for FullAdj. Based on the results in Section \ref{sec:simulation}, this indicates that spatial correlation is lost by not accounting for jittering. Additionally, $\sigma_\mathrm{S}^2$ is estimated higher for UnAdj and Smoothed than for FullAdj. This could be because the covariates are able to explain less of the spatial variation for the former two models.

For PopD and UrbR, there is a strong attenuation when jittering is ignored. The credible intervals for the coefficient of UrbR, $\beta_\mathrm{UrbR}$, suggest that $\beta_\mathrm{UrbR}$ is not significant at the 95\% level for UnAdj, whereas $\beta_\mathrm{UrbR}$ is  significant for FullAdj. This suggest that not accounting for jittering can lead to misleading conclusions. As discussed in Section \ref{sec:simulation}, it is hard to do direct comparisons in the estimated coefficients for Smoothed and for UnAdj and FullAdj since averaging covariates changes their meaning.

\begin{table}[H]
\centering
    \caption{Parameter estimates and the corresponding 95\% credible interval lengths in parentheses. Uncertainty is not computed for $\rho_\mathrm{S}$ and $\sigma_\mathrm{S}^2$.\label{tab:realData2columns}}
    \vspace{5mm}
\renewcommand{\belowrulesep}{0.05ex}
\begin{adjustbox}{max width=\textwidth}
    \begin{tabular}{ccccccccc}
       \toprule
        & \multicolumn{8}{c}{\textbf{Parameter}} \\[-5pt]
      \textbf{Model} & $\rho_\mathrm{S}$ & $\sigma_\mathrm{S}^{2}$ & $\mu$ & $\beta_\mathrm{DistW}$ & $\beta_\mathrm{CityA}$ & $\beta_\mathrm{Elev}$ & $\beta_\mathrm{PopD}$ & $\beta_\mathrm{UrbR}$ \\
      \midrule
\textbf{UnAdj} & ~64.15 & 1.91  & -2.24 \emph{(0.89)} & 0.91 \emph{(1.24)} &  -0.40 \emph{(0.13)} & -0.14 \emph{(0.69)} & 0.14 \emph{(0.09)} &  -0.14 \emph{(0.42)} \\
\textbf{Smoothed} & 58.42 & 1.88 & -2.32 \emph{(0.82)} & 1.08  \emph{(1.43)}  & -0.78 \emph{(0.29)}  & -0.40 \emph{(0.88)} & 0.49 \emph{(0.30)}  & -2.97 \emph{(2.20)}  \\
\textbf{FullAdj} & 107.68 & 1.65 & -2.21 \emph{(1.04)} & 0.62 \emph{(1.29)} & -0.43 \emph{(0.17)} & -0.02 \emph{(0.61)} &  0.32 \emph{(0.13)} & -1.35 \emph{(0.69)} \\
\bottomrule
\end{tabular}
\renewcommand{\belowrulesep}{0.65ex} 
\end{adjustbox}
\end{table}

\begin{figure}[H]
    \centering
    \subfigure[Predictions (UnAdj) \label{fig:predNN}]{\includegraphics[width=2.35in]{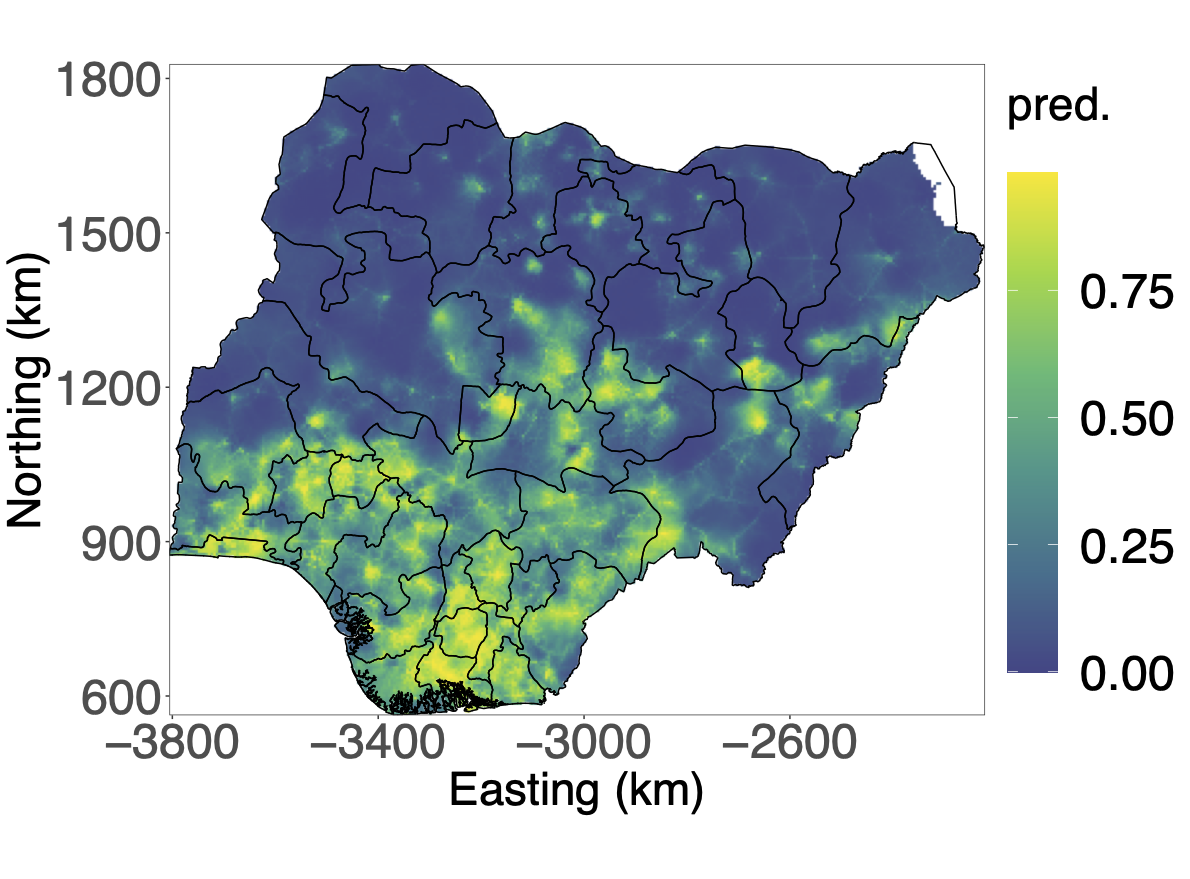}} 
    \subfigure[Uncertainty (UnAdj) \label{fig:uncertaintyNN}]{\includegraphics[width=2.35in]{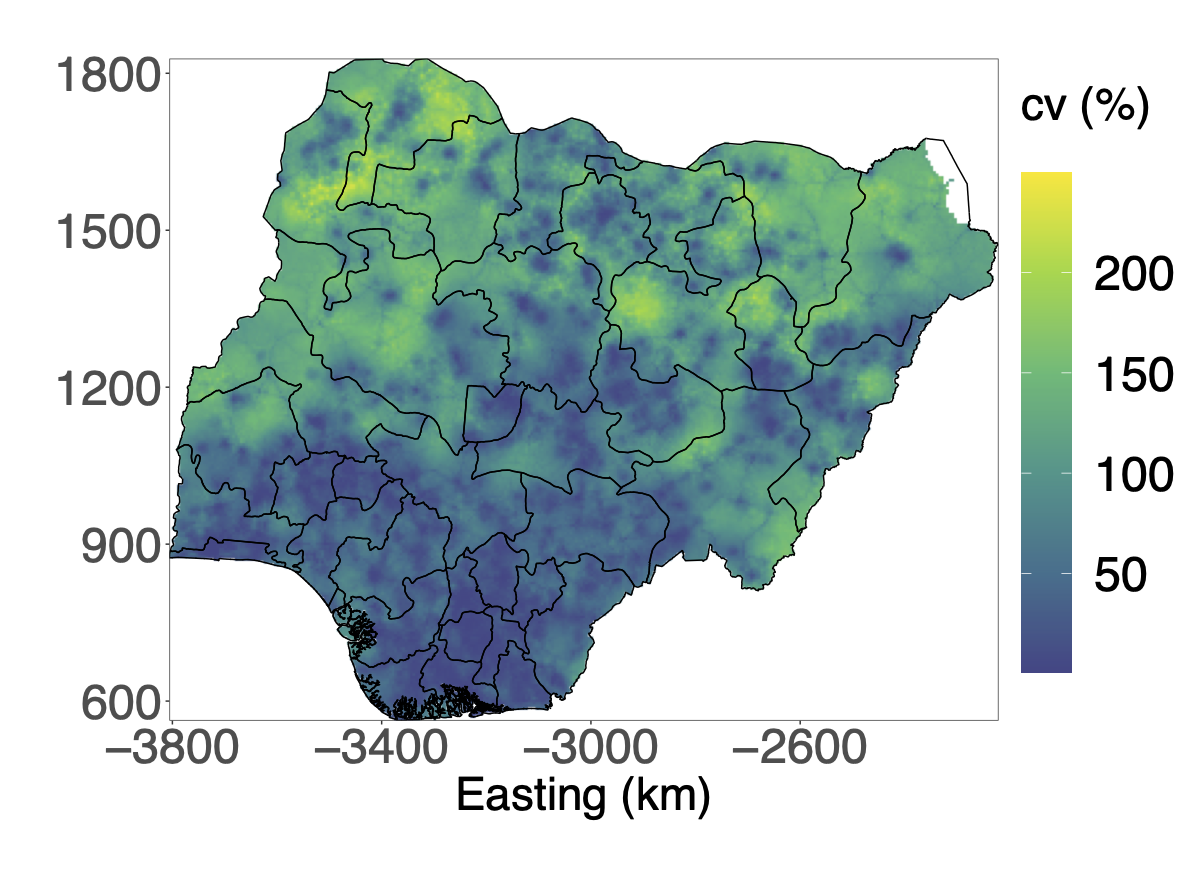}}
    \subfigure[Predictions (Smoothed) \label{fig:predNNsmoothed}]{\includegraphics[width=2.35in]{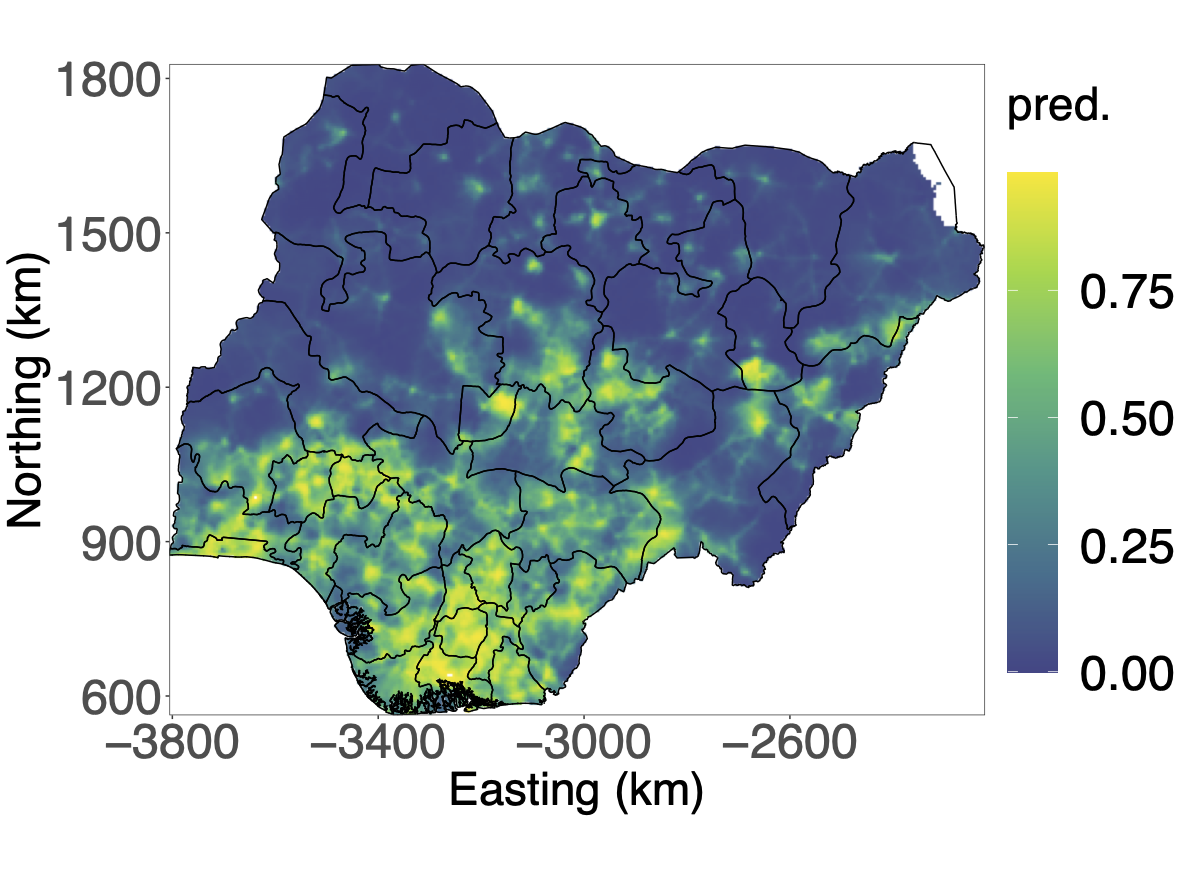}} 
    \subfigure[Uncertainty (Smoothed) \label{fig:uncertaintyNNsmoothed}]{\includegraphics[width=2.35in]{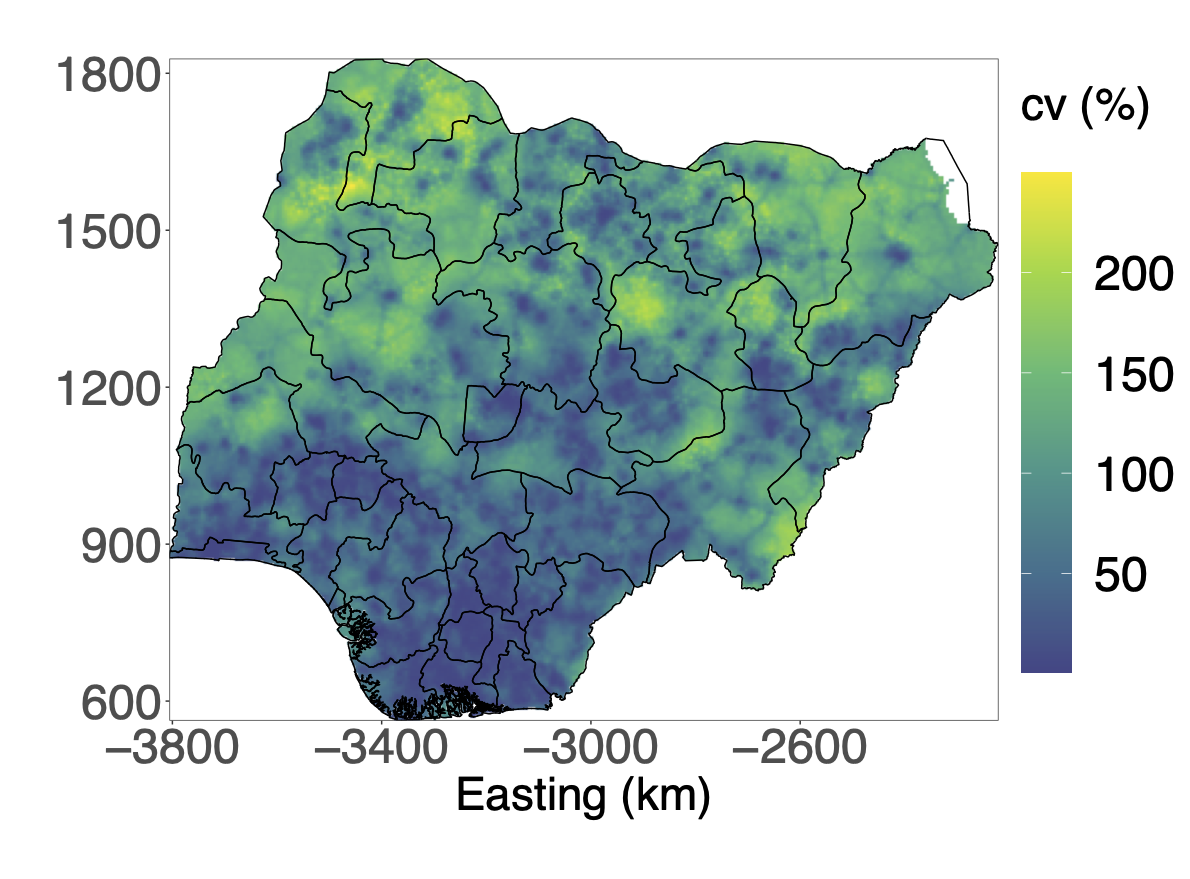}}
    \subfigure[Predictions (FullAdj)\label{fig:predCR}]{\includegraphics[width=2.35in]{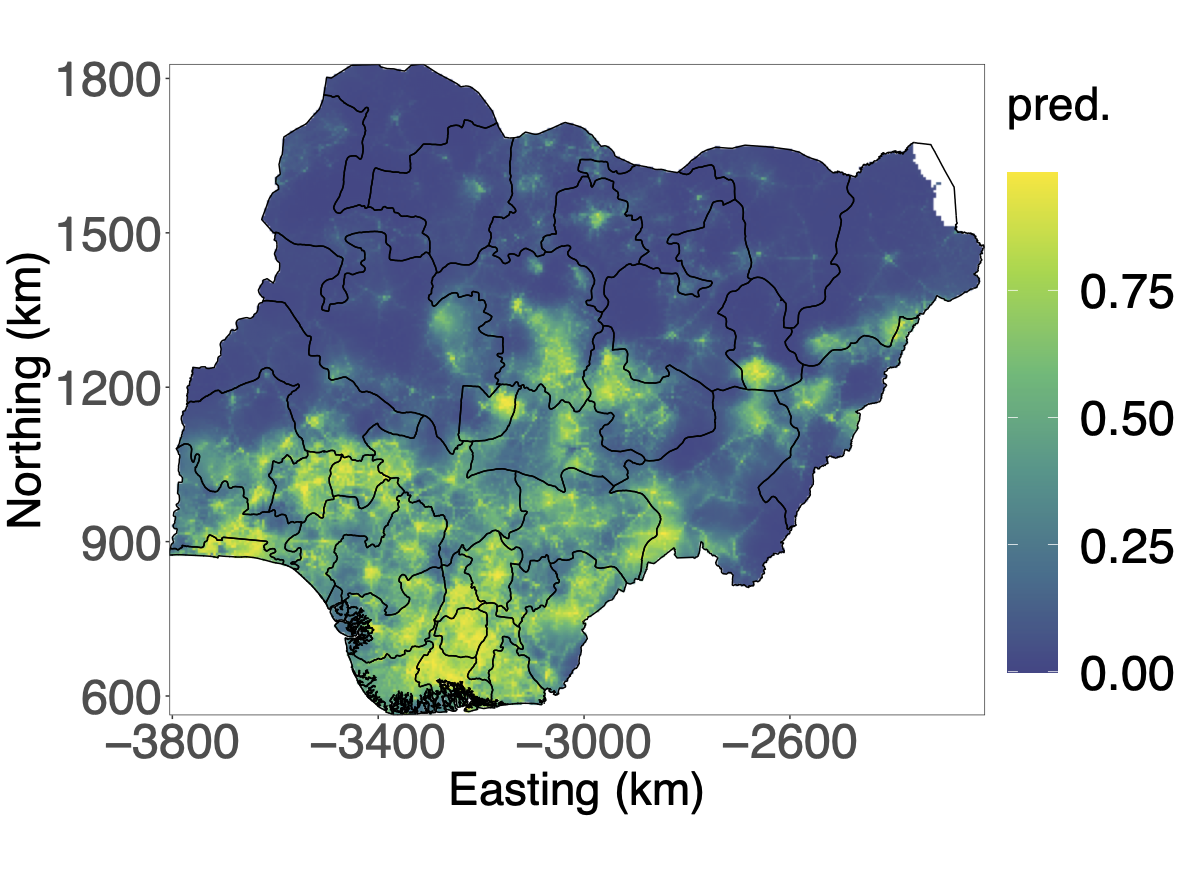}} 
    \subfigure[Uncertainty (FullAdj) \label{fig:uncertaintyCR}]{\includegraphics[width=2.35in]{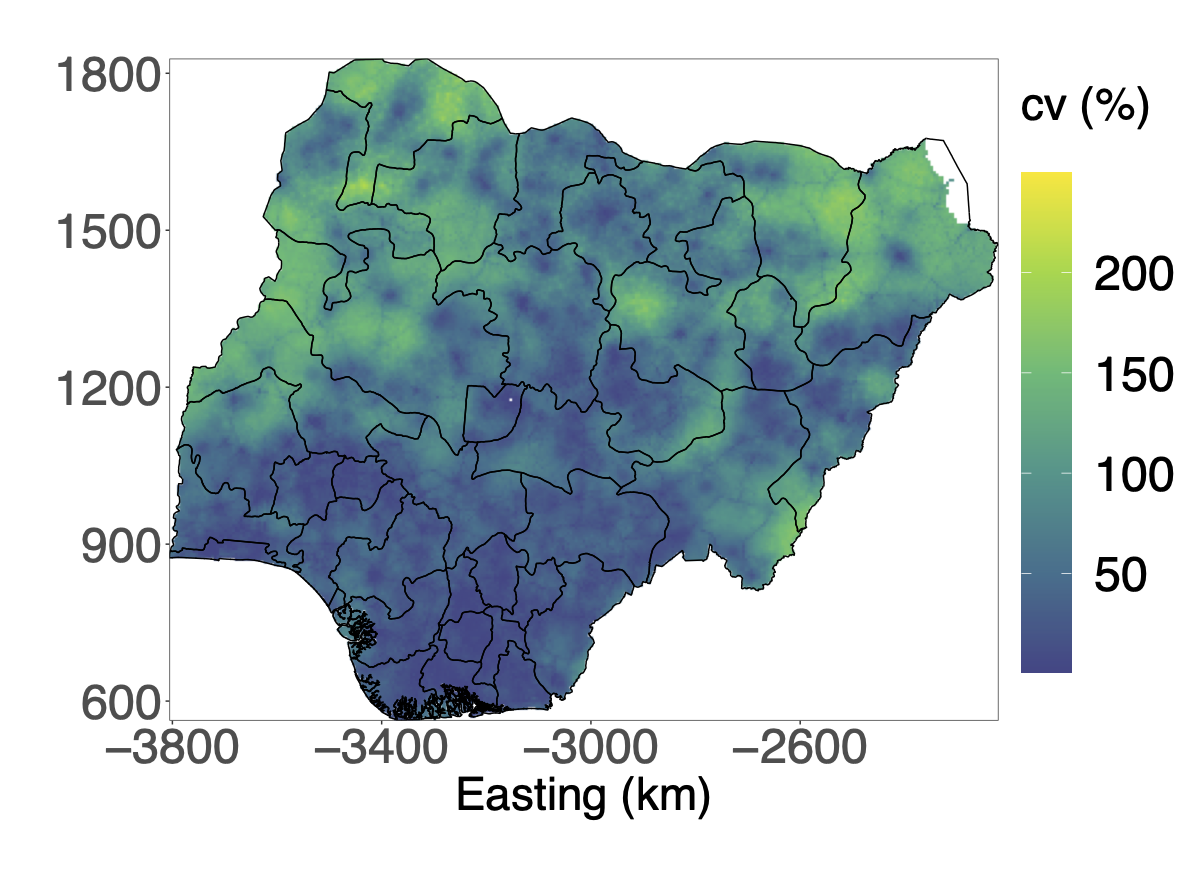}}
    \subfigure[Ratio of predictions \label{fig:ratioPred}]{\includegraphics[width=2.27in]{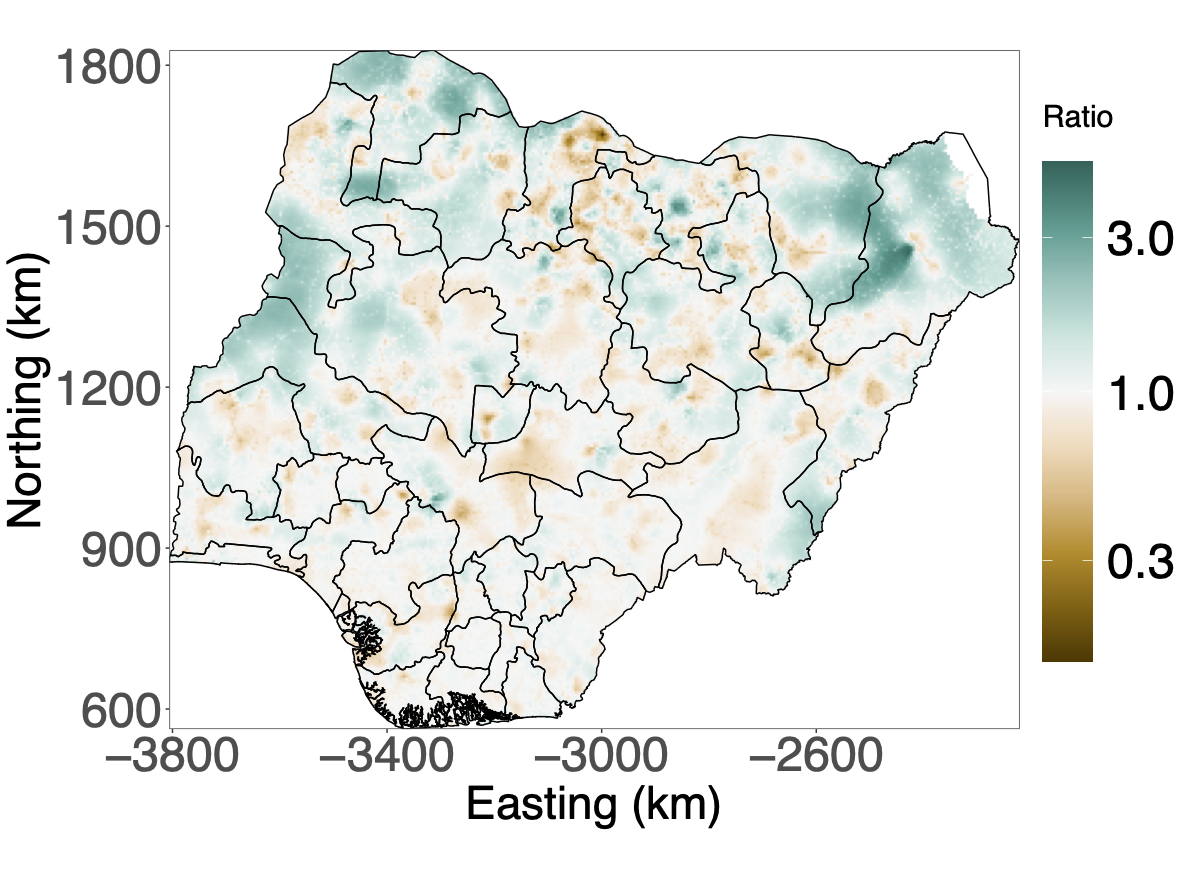}}
    \subfigure[Ratio of uncertainties \label{fig:ratioUncertainty}]{\includegraphics[width=2.27in]{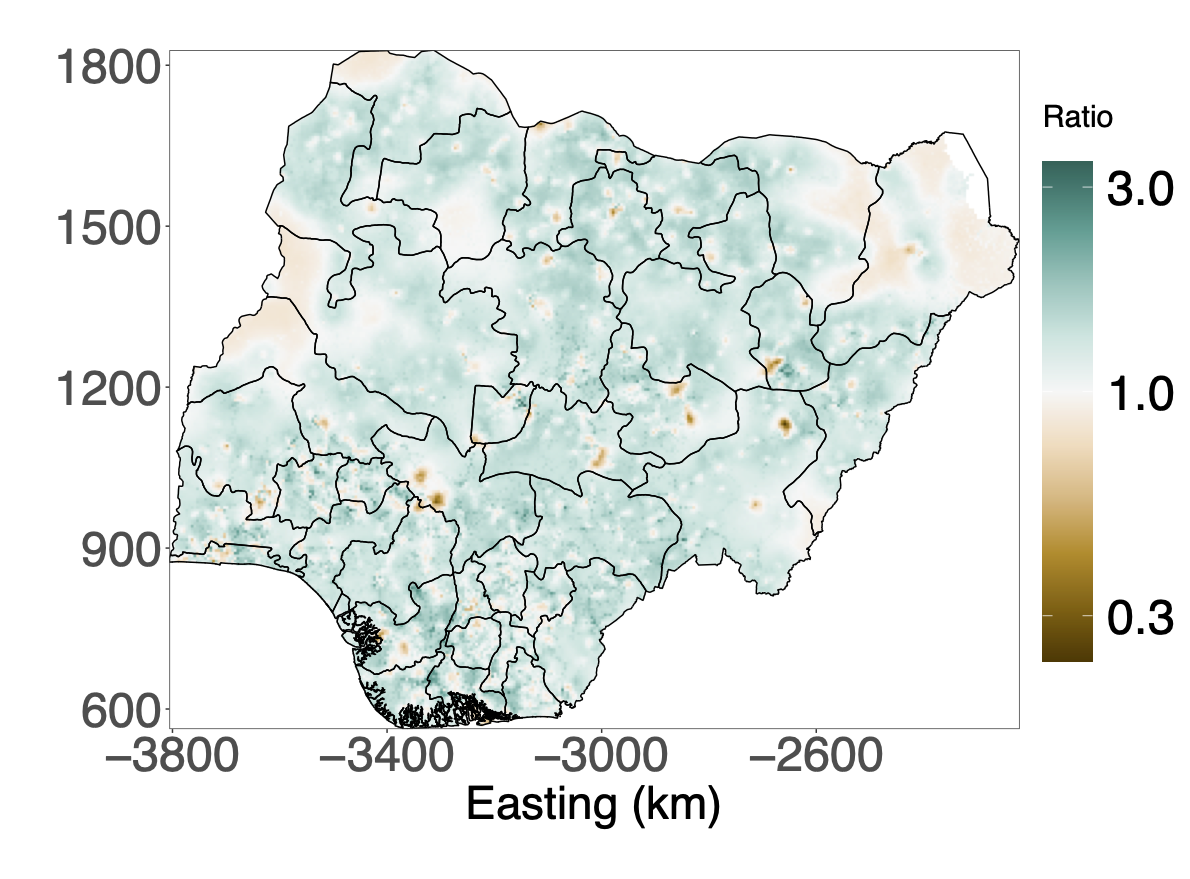}}
    
    \caption{Rows 1, 2 and 3 are predicted risk and the CVs for UnAdj, Smoothed, and FullAdj, respectively. Row 4 shows ratios (UnAdj/FullAdj) of predictions and CVs.}
    \label{fig:predEducation}
\end{figure}

Figure \ref{fig:predEducation} shows the $5\, \text{km}\times 5\, \text{km}$ pixel maps of predicted risk and CVs for UnAdj, Smoothed and FullAdj. The results show that some areas such as Borno (in the north-east) have up to three times the risk under the UnAdj approach as under FullAdj. Further, UnAdj tends to lead to higher uncertainty in the predictions than FullAdj.

\begin{figure}[H]
    \centering
    \subfigure[UnAdj \label{fig:areaPredNN}]{\includegraphics[width=2.35in]{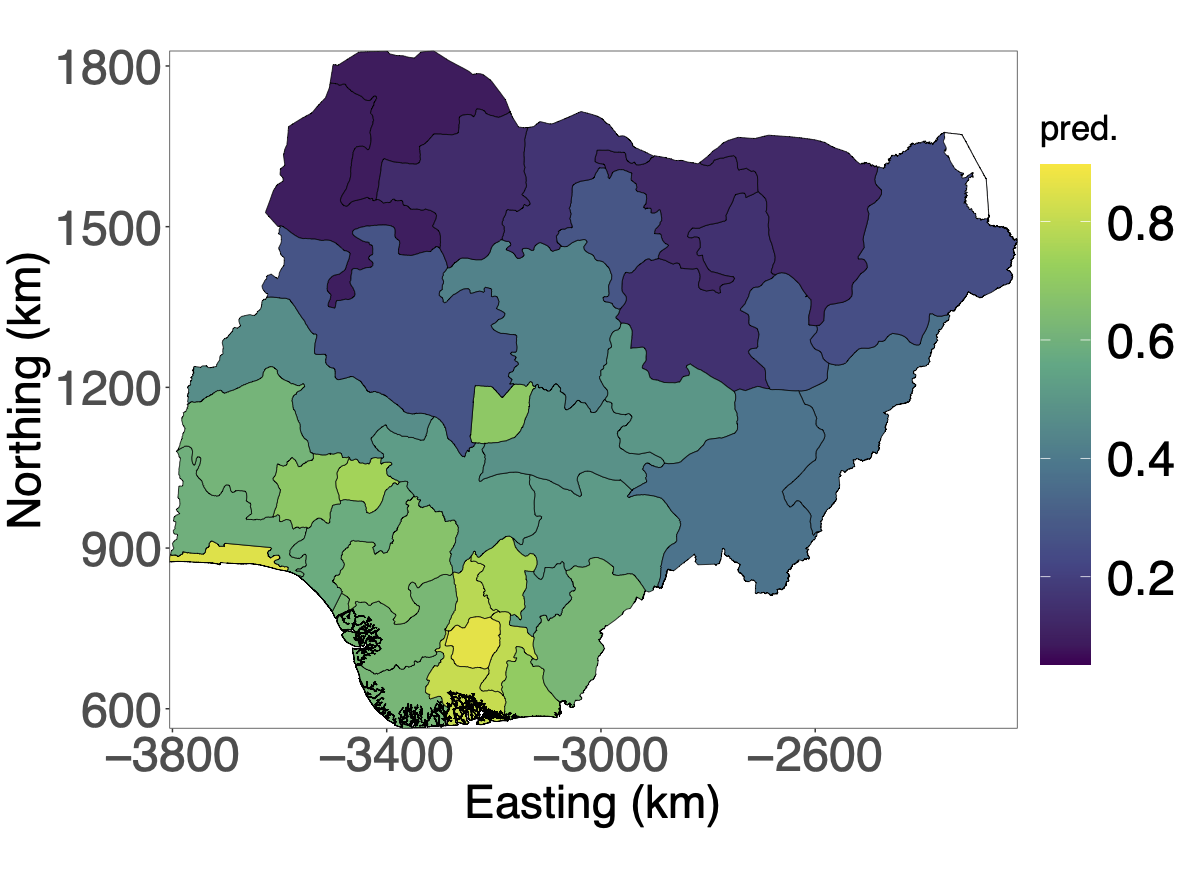}}
    \subfigure[UnAdj \label{fig:areaCVNN}]{\includegraphics[width=2.35in]{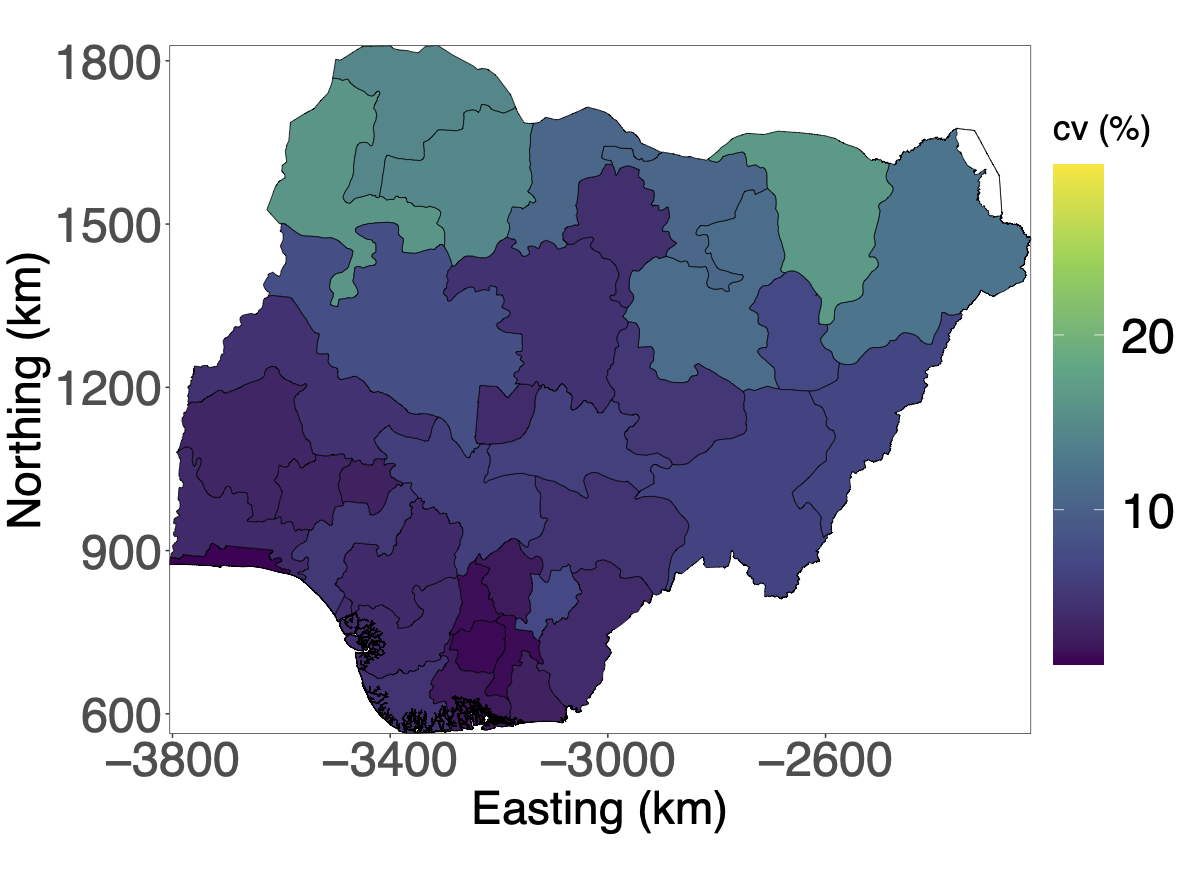}}
\subfigure[Smoothed \label{fig:areaPredNNsmoothed}]{\includegraphics[width=2.35in]{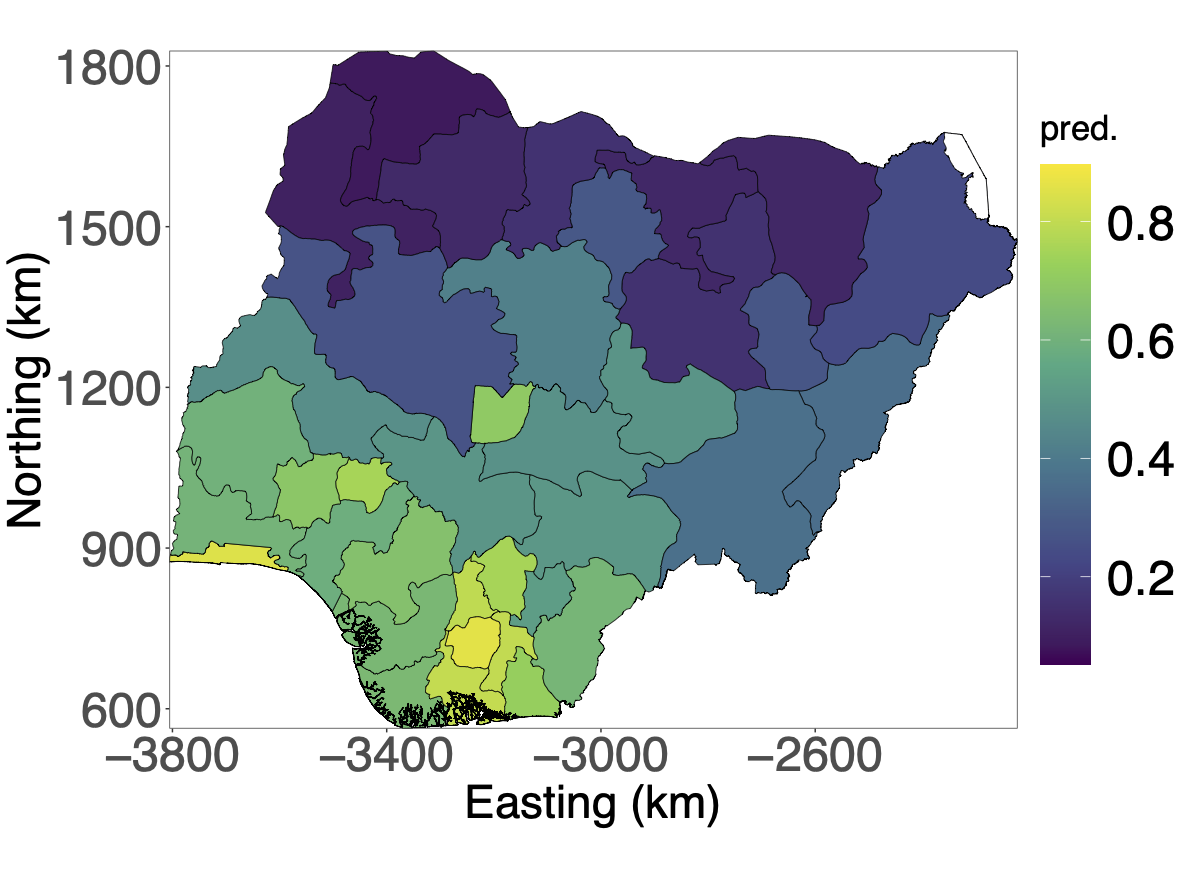}}
    \subfigure[Smoothed \label{fig:areaCVNNsmoothed}]{\includegraphics[width=2.35in]{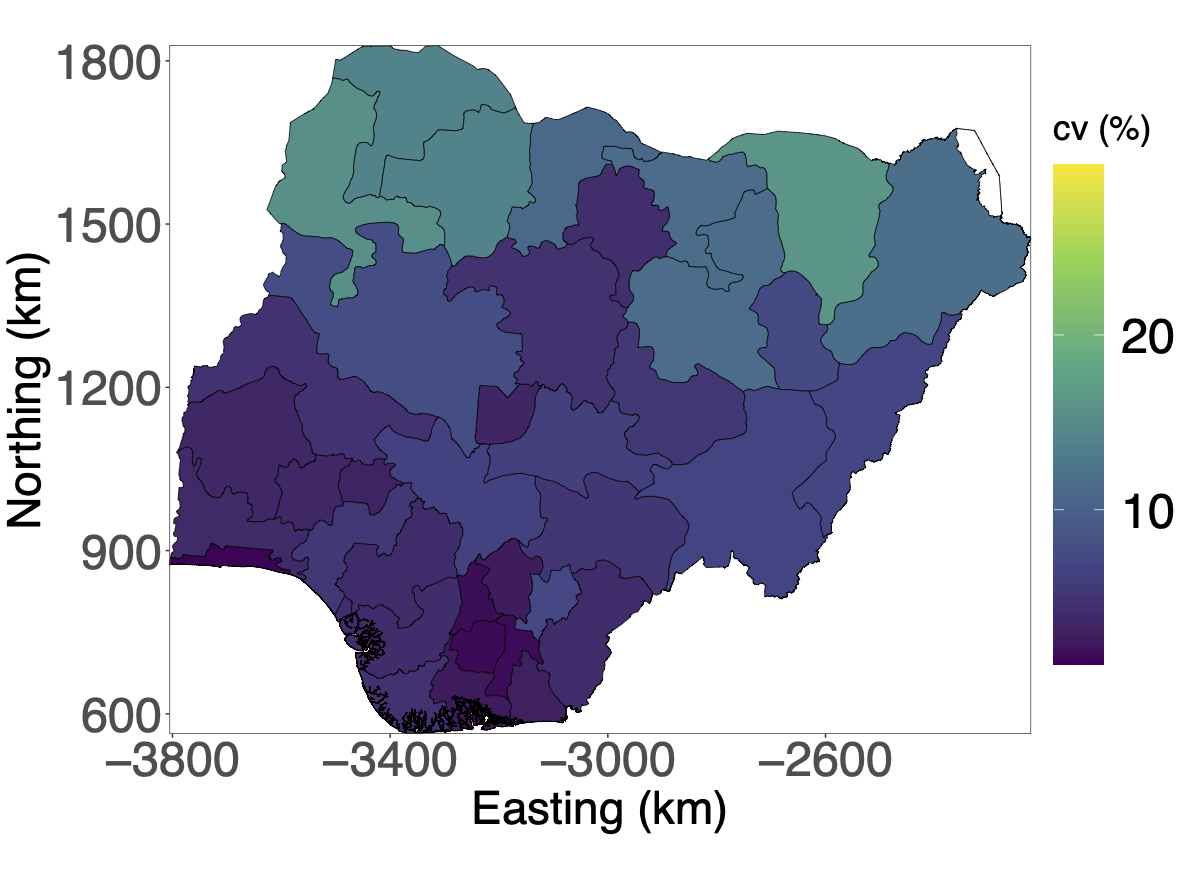}}
 \subfigure[FullAdj \label{fig:areaPredCR}]{\includegraphics[width=2.35in]{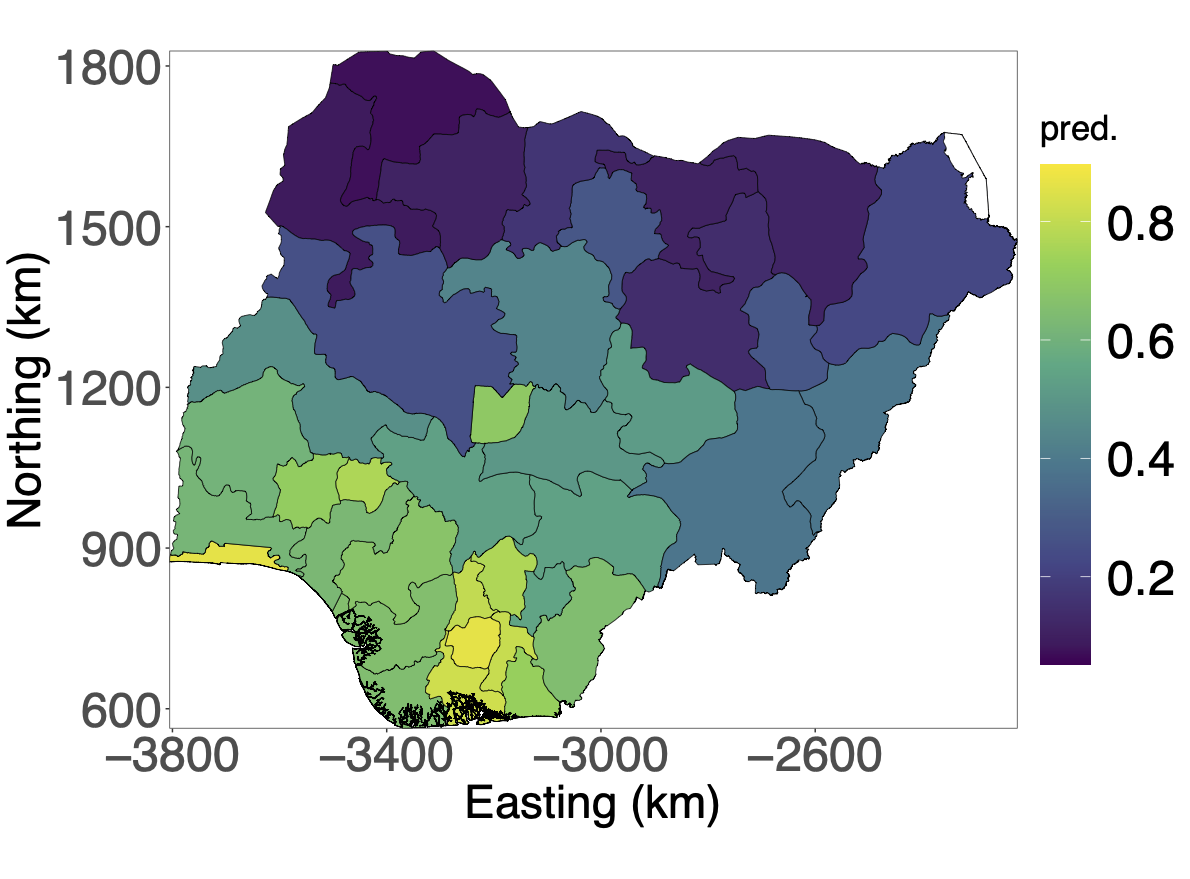}}
    \subfigure[FullAdj \label{fig:areaCVCR}]{\includegraphics[width=2.35in]{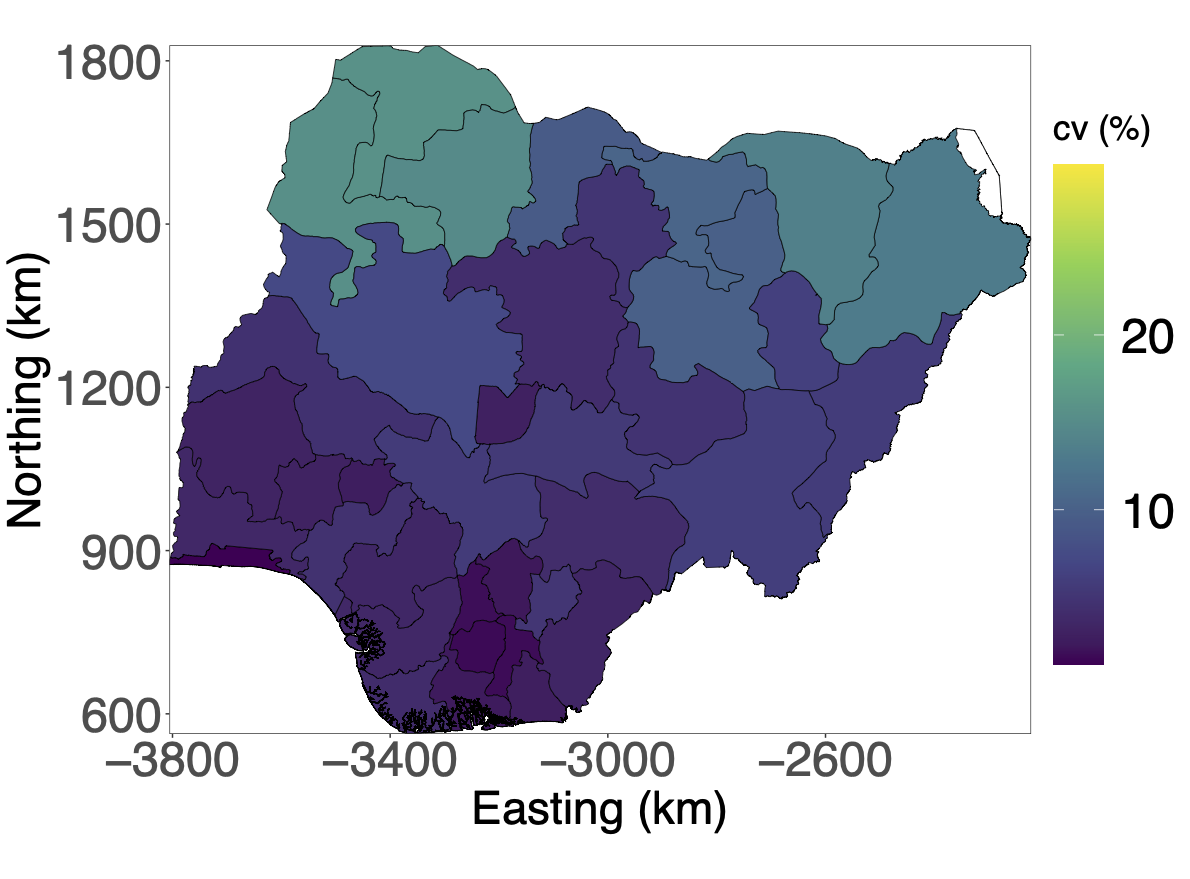}} 
    \subfigure[Ratio of aggregated predictions \label{fig:ratioAggregatedPred}]{\includegraphics[width=2.27in]{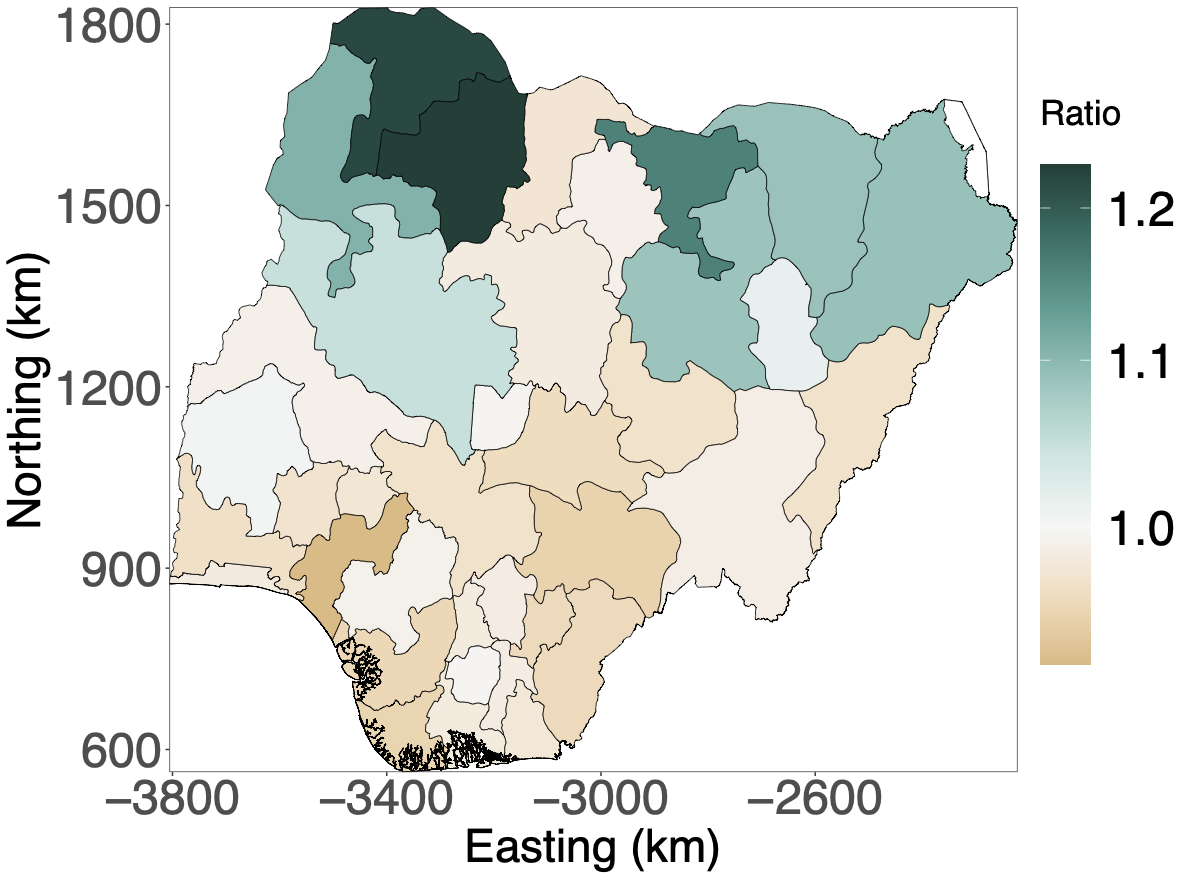}}
    \subfigure[Ratio of aggregated uncertainties \label{fig:ratioAggregatedUncertainty}]{\includegraphics[width=2.27in]{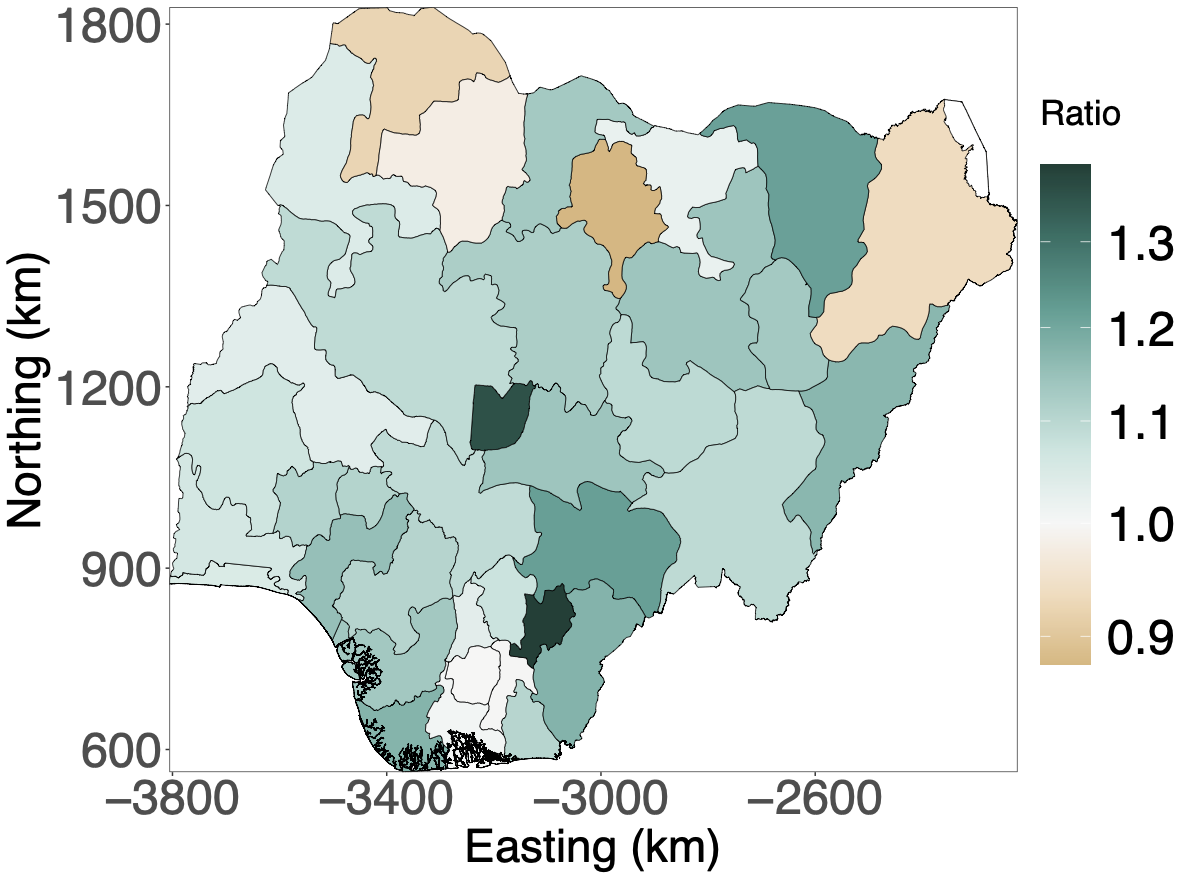}}
    \caption{Rows 1 and 2 are predicted risk and CVs for UnAdj and FullAdj respectively at the admin1 level, and row 3 shows ratios (UnAdj/FullAdj) of predictions and CVs.}
    \label{fig:area}
\end{figure}

We aggregate point level predictions with respect to population density to produce areal
estimates at the 37 admin1 areas (for more on aggregating point level predictions with respect to a population, see \citealt{paige2022spatial}). Figure \ref{fig:area} shows the predicted risk and associated CVs for UnAdj, Smoothed and FullAdj. From Figure \ref{fig:ratioAggregatedPred}, we see that the point estimates vary from a factor 0.9 to 1.2, and Figure \ref{fig:ratioAggregatedUncertainty} shows that some areas differ with a factor of up to 1.4 in CVs.

\begin{figure}[H]
    \centering
    \subfigure[Direct estimates \label{fig:meanVsMean}]{\includegraphics[width=2.35in]{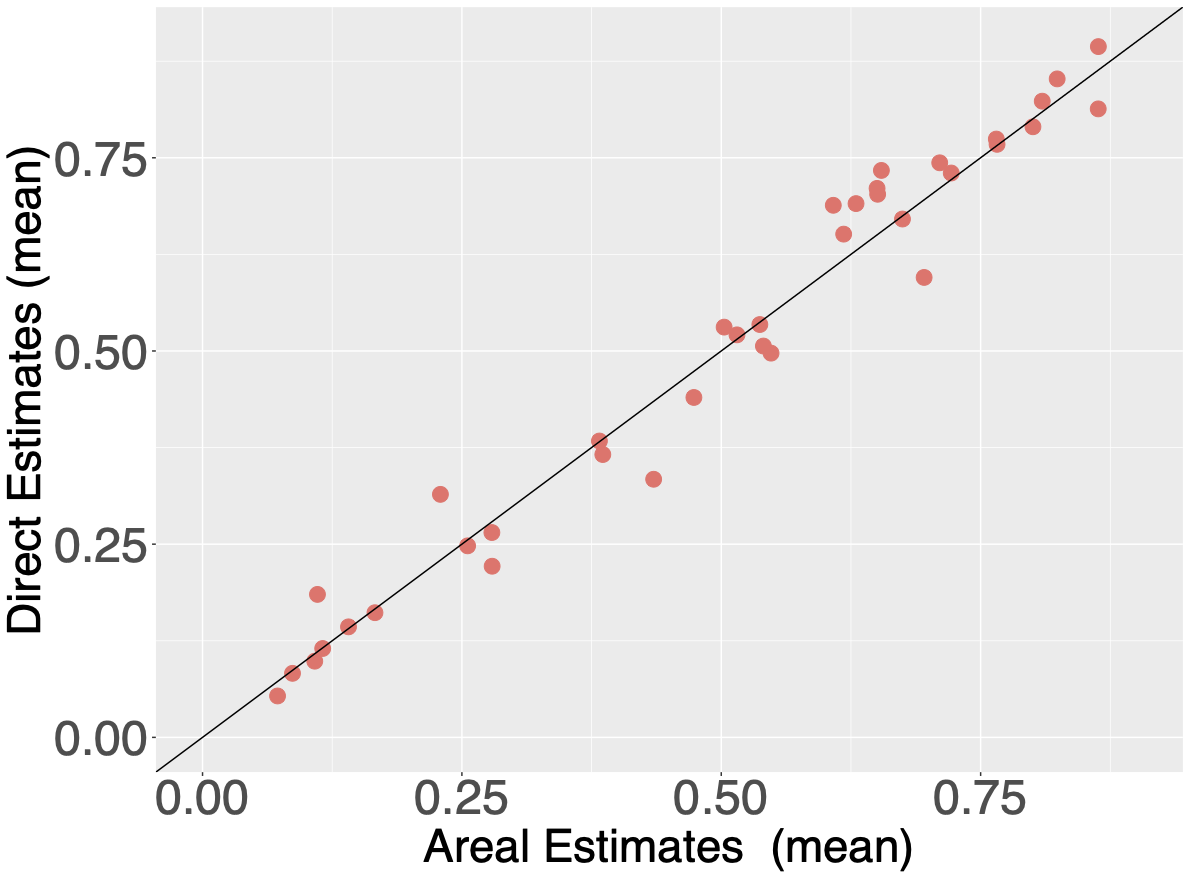}} 
    \subfigure[Uncertainty \label{fig:CVvsCV}]{\includegraphics[width=2.35in]{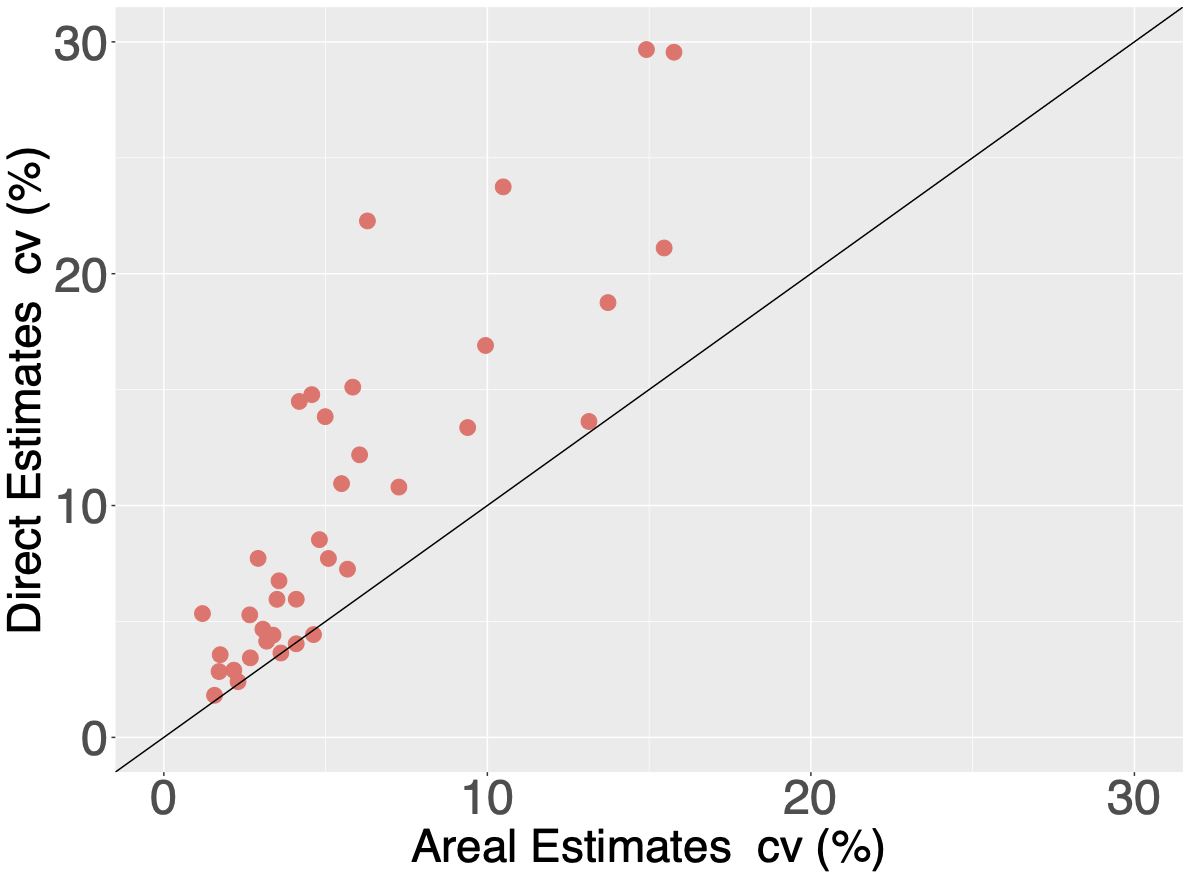}}
\caption{Comparison of the mean direct estimates against the areal means (left) and coefficient of variations (right) of UnAdj, Smoothed and FullAdj models.}
    \label{fig:DirectEstComparison1}
\end{figure}

In Figure \ref{fig:DirectEstComparison1}, we compare the areal estimates from the FullAdj model against direct estimates. Figure \ref{fig:CVvsCV} shows that the FullAdj reduces uncertainty compared to the direct estimates, and Figure \ref{fig:meanVsMean} shows that the estimates from FullAdj and the direct estimates are compatible with no unexpected deviations.

The ability to predict risk at unobserved
locations for UnAdj, CovAdj and FullAdj cannot 
be compared with cross validation. If data is
held-out from NDHS2018, we can only evaluate
the models' ability to predict risk at a new jittered cluster with unknown true location. However, the simulation study in
Section \ref{sec:simulation} indicates that FullAdj leads to an improvement in prediction if the data-generating model is the one estimated in this section. 

\section{Discussion \label{sec:discussion}}
Accounting for jittering substantially changed the parameter estimates for the geostatistical model for completion of secondary education among women aged 20--49 years. The simulation study demonstrated that these differences were linked to the strength of the signal of the spatial covariates when explaining the spatial variation. For strong signals, the associations were attenuated and the predictive power was reduced. 

The most important aspect of jittering in the context of geostatistical models for DHS data is to account for the resulting uncertainty in covariates extracted from rasters or extracted based on distances. This induces measurement error that may lead to attenuation in associations between covariates and the responses. Some covariates such as sanitation practices and household assets can be known exactly \citep{burgert2016guidance}, but these cannot be included when the goal is prediction since fine-scale rasters are not available.

In influential work such as \citet{burstein2019mapping} and \citet{local2021mapping}, covariates are resampled to a $5\,\text{km} \times 5\,\text{km}$ grid. This is similar, but slightly different than the Smoothed model discussed in this paper which uses $5\, \text{km}\times 5\, \text{km}$ windows around the observed locations. However, the results suggest that such approaches that average covariates do not address the issue of jittering, and do not improve predictions.

This paper used uniform priors for the unknown true locations. One could expect including information about population density and urbanicity into the priors would produce more accurate inference. However, population density maps and urbanicity maps are also modelled surfaces with biases and uncertainties that are not well understood. This means that evaluation of the sensitivity to such maps would have to be investigated, and one would need a way to evaluate whether such a model works better.

The inference scheme uses empirical Bayes. It is possible to investigate methods such as
INLA, but the implementation in the R \citep{Rsoftware} package \texttt{inla} does not allow the likelihood to depend on the latent risk at multiple locations, which is necessary due to the integration points. MCMC algorithms such as STAN \citep{stan2020} has the required flexibility, but is infeasible for thousands of spatial locations.

When analysing completion of secondary education, we found that, for urbanicity, an effect size of 0 was contained in the 95\% credible interval when not adjusting for jittering, and not contained when adjusting for jittering. This suggests that not accounting for jittering when analysing DHS data is a practice that can alter conclusions about statistical significance. Since the proposed approach is fast for spatial analysis, we suggest to use the new approach routinely for analysing DHS data to avoid the risk of misleading conclusions and reduced predictive power. The code used throughout the paper can be found in the GitHub repository \url{https://github.com/umut-altay/GeoAdjust}.

\bibliography{smj-template}

\begin{appendices}

\section{Introduction}
This document contains supplementary results for the paper titled "Impact of Jittering on Raster- and Distance-based Geostatistical Analysis of DHS Data". Section \ref{sec:2} presents results of a version of the simulation in the main paper where the true locations are fixed to match the design of the Nigeria 2018 DHS survey. Section \ref{sec:3} consists of the supplementary results for the main simulation study where the coordinates of the true locations are sampled according population density.

\section{Simulation results with fixed true locations \label{sec:2}}

\begin{figure}[H]
\centering
\includegraphics[width=\textwidth]{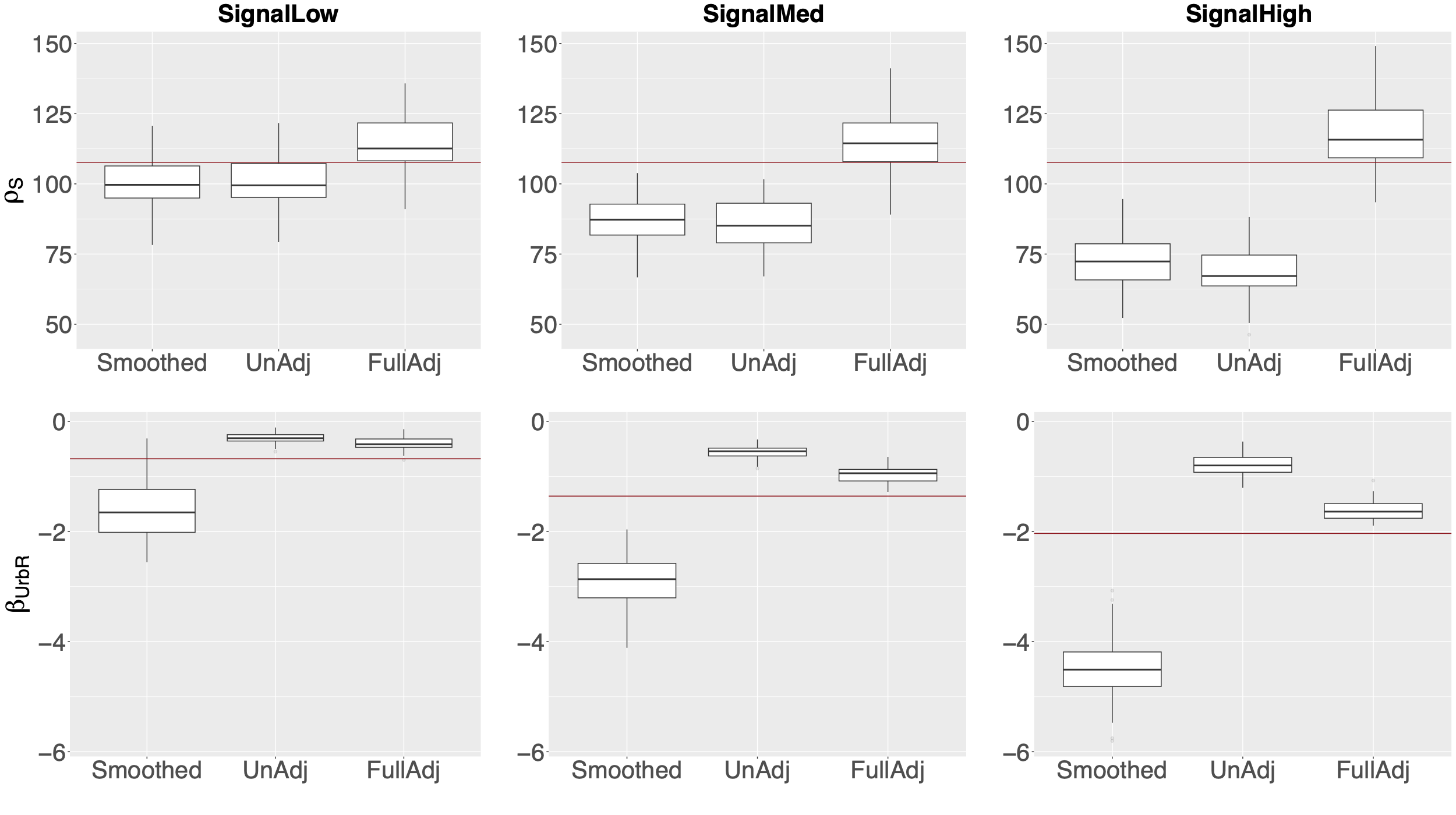}
\caption{Box plots of estimated $\rho_\mathrm{S}$ and $\beta_{\mathrm{UrbR}}$ for SignalLow, SignalMed and SignalHigh. The horizontal red lines show the true parameter value.}
\label{fig:urbanRangeFixedLocations}
\end{figure}

\begin{figure}[H]
\centering
\includegraphics[width=\textwidth]{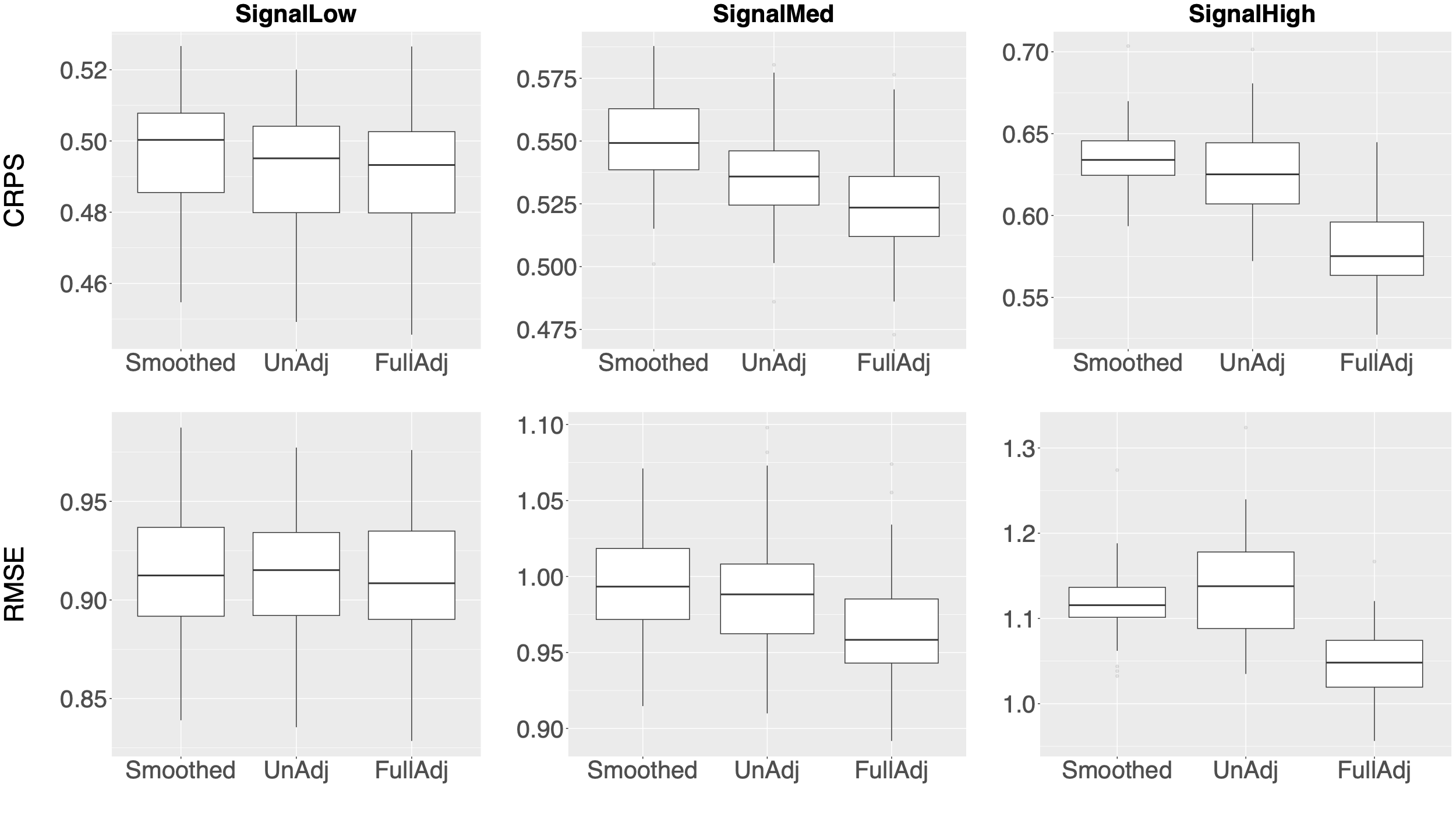}
\caption{Box plots of CRPS and RMSE for predictions for SignalLow, SignalMed and SignalHigh.}
   \label{fig:scoresFixedLocations}
\end{figure}

Figure \ref{fig:urbanRangeFixedLocations} shows the box plots of the spatial range and the urbanization coefficient estimates with Smoothed, UnAdj and FullAdj models for low, medium and high signal strength levels, respectively. Figure \ref{fig:scoresFixedLocations} shows the box plots of CRPS and RMSE for predictions with Smoothed, UnAdj and FullAdj models for low, medium and high signal strength levels, respectively.

Tables \ref{tab:scaledHalffixedLocs}, \ref{tab:scaled1fixedLocs} and \ref{tab:scaled1point5fixedLocs} contain average bias and RMSE values of the model parameter estimates together with the predictive performance measures for Smoothed, UnAdj and FullAdj models for low, medium and high signal strength levels, respectively.

\begin{table}[H]
\centering
    \caption{Average bias and RMSE (in parantheses) of model parameter estimates, together with the average predictive measures (RMSE and CRPS) for UnAdj, Smoothed, and FullAdj models. The results belong to the signal strength level "SignalLow". \label{tab:scaledHalffixedLocs}}
    \vspace{5mm}
    \begin{adjustbox}{max width=\textwidth}
    \begin{tabular}{|c|cc|c|}
   \toprule\toprule
    & \multicolumn{2}{c|}{\textbf{Unadjusted}} & \textbf{Adjusted} \\
      \midrule
         \textbf{Model} & \textbf{UnAdj} & \textbf{Smoothed} & \textbf{FullAdj} \\
        \midrule
        \textbf{Parameter} &  &  &  \\
 \midrule
        ~~$\rho$               & -7.33 \emph{(12.08)} & -7.10  \emph{(11.75)} &  6.47 \emph{(12.19)}    \\
        ~~$\sigma^{2}$           & 0.04 \emph{(0.09)} & 0.04    \emph{(0.10)} &   0.03 \emph{(0.09)}    \\
        ~~$\mu$                  & 0.15 \emph{(0.28)} & 0.10    \emph{(0.25)} &   0.17 \emph{(0.29)}     \\ 
        ~~$\beta_\mathrm{dist}$  & -0.08 \emph{(0.37)} & -0.03  \emph{(0.45)} &   -0.09 \emph{(0.36)}    \\ 
        ~~$\beta_\mathrm{tTime}$ & 0.03 \emph{(0.04)} & -0.13   \emph{(0.15)} &    0.03\emph{(0.04)}     \\ 
        ~~$\beta_\mathrm{elev}$  & 0.01 \emph{(0.17)} & -0.0001 \emph{(0.19)} &    0.009\emph{(0.14)}     \\ 
        ~~$\beta_\mathrm{pop}$   & -0.06 \emph{(0.07)} & 0.10   \emph{(0.13)} &   -0.05 \emph{(0.05)}     \\ 
        ~~$\beta_\mathrm{urb}$   & 0.37 \emph{(0.38)} & -0.92   \emph{(1.05)} &   0.27\emph{(0.30)}    \\ 
        \midrule
     \textbf{Predictive measures} &  &  &  \\
        \midrule
        ~~RMSE & 0.91 & 0.91 & 0.91 \\ 
        ~~CRPS & 0.49 & 0.49 & 0.49 \\ 
        
 \bottomrule
\end{tabular}
\end{adjustbox}
\end{table}

\begin{table}[H]
\centering
    \caption{Average bias and RMSE (in parantheses) of model parameter estimates, together with the average predictive measures (RMSE and CRPS) for UnAdj, Smoothed, and FullAdj models. The results belong to the signal strength level "SignalMed".  \label{tab:scaled1fixedLocs}}
    \vspace{5mm}
    \begin{adjustbox}{max width=\textwidth}
    \begin{tabular}{|c|cc|c|}
   \toprule\toprule
    & \multicolumn{2}{c|}{\textbf{Unadjusted}} & \textbf{Adjusted} \\
      \midrule
         \textbf{Model} & \textbf{UnAdj} & \textbf{Smoothed} & \textbf{FullAdj} \\
        \midrule
        \textbf{Parameter} &  &  & \\
\midrule
        ~~$\rho$                 & -22.03\emph{(23.70)} & -20.39\emph{(22.02)} &  7.96 \emph{(13.44)}    \\
        ~~$\sigma^{2}$           &  0.02 \emph{(0.08)} & 0.04  \emph{(0.09)} &  -0.01 \emph{(0.09)}    \\
        ~~$\mu$                  &  0.34 \emph{(0.42)} & 0.22  \emph{(0.33)} &  0.35 \emph{(0.42)}     \\ 
        ~~$\beta_\mathrm{dist}$  &  -0.05\emph{(0.33)} & 0.06  \emph{(0.42)} &  -0.09 \emph{(0.34)}     \\ 
        ~~$\beta_\mathrm{tTime}$ &  0.08 \emph{(0.09)} & -0.25 \emph{(0.26)} &  0.04 \emph{(0.05)}     \\ 
        ~~$\beta_\mathrm{elev}$  &  0.01 \emph{(0.19)} & -0.02 \emph{(0.22)} &  0.02 \emph{(0.16)}     \\ 
        ~~$\beta_\mathrm{pop}$   &  -0.12\emph{(0.13)} & 0.22  \emph{(0.24)} &  -0.08 \emph{(0.08)}     \\ 
        ~~$\beta_\mathrm{urb}$   &  0.78 \emph{(0.79)} & -1.56 \emph{(1.64)} &  0.40 \emph{(0.42)}    \\ 
        \midrule
     \textbf{Predictive measures} &  &  &  \\
        \midrule
        ~~RMSE & 0.99 & 0.99  & 0.96 \\ 
        ~~CRPS & 0.53 & 0.55  & 0.52 \\ 
        
 \bottomrule
\end{tabular}
\end{adjustbox}
\end{table}

\begin{table}[H]
\centering
    \caption{Average bias and RMSE (in parantheses) of model parameter estimates, together with the average predictive measures (RMSE and CRPS) for UnAdj, Smoothed, and FullAdj models. The results belong to the signal strength level "SignalHigh".   \label{tab:scaled1point5fixedLocs}}
    \vspace{5mm}
    \begin{adjustbox}{max width=\textwidth}
    \begin{tabular}{|c|cc|c|}
   \toprule\toprule
    & \multicolumn{2}{c|}{\textbf{Unadjusted}} & \textbf{Adjusted} \\
      \midrule
         & \textbf{UnAdj} & \textbf{Smoothed} & \textbf{FullAdj} \\
        \midrule
        \textbf{Parameter} &  &  &  \\
 \midrule
        ~~$\rho$                 & -39.07 \emph{(40.10)} & -35.16 \emph{(36.27)}&      9.52 \emph{(15.44)}    \\
        ~~$\sigma^{2}$           &  0.13\emph{(0.15)}    & 0.15   \emph{(0.17)} &      0.007 \emph{(0.11)}    \\
        ~~$\mu$                  &  0.48\emph{(0.53)}    & 0.27   \emph{(0.36)} &      0.40 \emph{(0.47)}     \\ 
        ~~$\beta_\mathrm{dist}$  & -0.10\emph{(0.37)}    & 0.01   \emph{(0.43)} &     -0.12\emph{(0.37)}     \\ 
        ~~$\beta_\mathrm{tTime}$ &  0.13\emph{(0.14)}    & -0.37  \emph{(0.38)} &      0.05\emph{(0.06)}     \\ 
        ~~$\beta_\mathrm{elev}$  & 0.002\emph{(0.17)}    & -0.006 \emph{(0.25)} &     -0.002\emph{(0.15)}     \\ 
        ~~$\beta_\mathrm{pop}$   & -0.22 \emph{(0.22)}   & 0.31   \emph{(0.32)} &     -0.10\emph{(0.11)}     \\ 
        ~~$\beta_\mathrm{urb}$   & 1.24 \emph{(1.25)}    & -2.43  \emph{(2.49)} &      0.41 \emph{(0.45)}    \\ 
        \midrule
     \textbf{Predictive measures} &  &  &  \\
        \midrule
        ~~RMSE & 1.13 & 1.11  & 1.08 \\ 
        ~~CRPS & 0.62 & 0.63  & 0.75 \\  
        
 \bottomrule
\end{tabular}
\end{adjustbox}
\end{table}

\section{Additional results for randomly generated true locations \label{sec:3}}

Tables \ref{tab:scaledHalfgeneratedLocs}, \ref{tab:scaled1generatedLocs} and \ref{tab:scaled1point5generatedLocs} contain average bias and RMSE values of the model parameter estimates together with the predictive performance measures for Smoothed, UnAdj and FullAdj models for low, medium and high signal strength levels, respectively.

\begin{table}[H]
\centering
    \caption{Average bias and RMSE (in parantheses) of model parameter estimates, together with the average predictive measures (RMSE and CRPS) for UnAdj, Smoothed, and FullAdj models. The results belong to the signal strength level "SignalLow". \label{tab:scaledHalfgeneratedLocs}}
    \vspace{5mm}
    \begin{adjustbox}{max width=\textwidth}
    \begin{tabular}{|c|cc|c|}
   \toprule\toprule
    & \multicolumn{2}{c|}{\textbf{Unadjusted}} & \textbf{Adjusted}\\
      \midrule
         \textbf{Model} & \textbf{UnAdj} & \textbf{Smoothed} & \textbf{FullAdj} \\
        \midrule
        \textbf{Parameter} &  &  & \\
  \midrule
        ~~$\rho$                 & -7.40 \emph{(12.01)} & -6.92 \emph{(11.70)} & 6.67 \emph{(12.48)}    \\
        ~~$\sigma^{2}$           & 0.04 \emph{(0.09)} & 0.04   \emph{(0.09)} & 0.03  \emph{(0.10)}    \\
        ~~$\mu$                  & 0.21 \emph{(0.33)} & 0.15   \emph{(0.30)} &  0.21 \emph{(0.34)}    \\ 
        ~~$\beta_\mathrm{dist}$  & -0.10 \emph{(0.35)} & -0.04 \emph{(0.42)} & -0.11 \emph{(0.36)}    \\  
        ~~$\beta_\mathrm{tTime}$ & 0.03 \emph{(0.04)} & -0.13  \emph{(0.15)} &  0.02 \emph{(0.03)}    \\  
        ~~$\beta_\mathrm{elev}$  & -0.003 \emph{(0.13)} & -0.01\emph{(0.22)} &  0.01 \emph{(0.13)}    \\ 
        ~~$\beta_\mathrm{pop}$   & -0.08 \emph{(0.08)} & 0.07  \emph{(0.10)} &  -0.05 \emph{(0.06)}    \\ 
        ~~$\beta_\mathrm{urb}$   & 0.39 \emph{(0.41)} & -0.49  \emph{(0.70)} &  0.23 \emph{(0.27)}    \\ 
        \midrule
     \textbf{Predictive measures} &  &  & \\
        \midrule
        ~~RMSE & 0.93 & 0.93 & 0.92 \\ 
        ~~CRPS & 0.50 & 0.50 & 0.49 \\  
        
 \bottomrule
\end{tabular}
\end{adjustbox}
\end{table}

\begin{table}[H]
\centering
    \caption{Average bias and RMSE (in parantheses) of model parameter estimates, together with the average predictive measures (RMSE and CRPS) for UnAdj, Smoothed, and FullAdj models. The results belong to the signal strength level "SignalMed".  \label{tab:scaled1generatedLocs}}
    \vspace{5mm}
    \begin{adjustbox}{max width=\textwidth}
    \begin{tabular}{|c|cc|c|}
   \toprule\toprule
    & \multicolumn{2}{c|}{\textbf{Unadjusted}} & \textbf{Adjusted} \\
      \midrule
         \textbf{Model} & \textbf{UnAdj} & \textbf{Smoothed} & \textbf{FullAdj} \\
        \midrule
        \textbf{Parameter} &  &  &  \\
 \midrule
        ~~$\rho$                 & -21.32 \emph{(22.48)} & -19.97 \emph{(21.37)}  & 7.51 \emph{(11.96)}    \\
        ~~$\sigma^{2}$           & 0.05 \emph{(0.09)} & 0.05  \emph{(0.09)} & 0.009 \emph{(0.09)}    \\
        ~~$\mu$                  & 0.32 \emph{(0.42)} & 0.21 \emph{(0.34)} &  0.27 \emph{(0.37)}    \\ 
        ~~$\beta_\mathrm{dist}$  & -0.0001 \emph{(0.27)} & 0.14 \emph{(0.37)} & 0.006 \emph{(0.26)}    \\  
        ~~$\beta_\mathrm{tTime}$ & 0.08 \emph{(0.09)} & -0.22 \emph{(0.24)} & 0.02 \emph{(0.04)}    \\  
        ~~$\beta_\mathrm{elev}$  & 0.01 \emph{(0.19)} & 0.04 \emph{(0.29)} & 0.02  \emph{(0.17)}    \\ 
        ~~$\beta_\mathrm{pop}$   & -0.16 \emph{(0.16)} & 0.15 \emph{(0.17)} & -0.09  \emph{(0.09)}    \\ 
        ~~$\beta_\mathrm{urb}$   & 0.84 \emph{(0.85)} & -1.03 \emph{(1.17)} & 0.37 \emph{(0.42)}    \\ 
        \midrule
     \textbf{Predictive measures} &  &  & \\
        \midrule
        ~~RMSE & 1.02 & 1.02 & 0.99  \\ 
        ~~CRPS & 0.55 & 0.57 & 0.54  \\  
        
 \bottomrule
\end{tabular}
\end{adjustbox}
\end{table}

\begin{table}[H]
\centering
    \caption{Average bias and RMSE (in parantheses) of model parameter estimates, together with the average predictive measures (RMSE and CRPS) for UnAdj, Smoothed, and FullAdj models. The results belong to the signal strength level "SignalHigh".  \label{tab:scaled1point5generatedLocs}}
    \vspace{5mm}
    \begin{adjustbox}{max width=\textwidth}
    \begin{tabular}{|c|cc|c|}
   \toprule\toprule
    & \multicolumn{2}{c|}{\textbf{Unadjusted}} & \textbf{Adjusted} \\
      \midrule
         & \textbf{UnAdj} & \textbf{Smoothed} &  \textbf{FullAdj} \\
        \midrule
        \textbf{Parameter} &  &  & \\
 \midrule
        ~~$\rho$                 &  -41.45 \emph{(42.23)} & -38.81 \emph{(39.80)} & 6.18 \emph{(14.29)}    \\
        ~~$\sigma^{2}$           & 0.12 \emph{(0.15)}  & 0.11 \emph{(0.14)} &  -0.03 \emph{(0.11)}    \\
        ~~$\mu$                  & 0.47 \emph{(0.52)} & 0.25 \emph{(0.34)} &  0.35 \emph{(0.42)}    \\ 
        ~~$\beta_\mathrm{dist}$  & -0.03 \emph{(0.32)} & 0.15  \emph{(0.44)} & -0.04  \emph{(0.33)}    \\  
        ~~$\beta_\mathrm{tTime}$ & 0.13 \emph{(0.13)} & -0.38  \emph{(0.39)} & 0.02 \emph{(0.05)}    \\  
        ~~$\beta_\mathrm{elev}$  & 0.03 \emph{(0.21)} & 0.06 \emph{(0.33)} & 0.03  \emph{(0.21)}    \\ 
        ~~$\beta_\mathrm{pop}$   & -0.26 \emph{(0.26)} & 0.17 \emph{(0.20)} &  -0.11 \emph{(0.11)}    \\ 
        ~~$\beta_\mathrm{urb}$   & 1.25 \emph{(1.26)} & -1.44 \emph{(1.62)} & 0.41  \emph{(0.45)}    \\ 
        \midrule
     \textbf{Predictive measures} &  &  & \\
        \midrule
        ~~RMSE & 1.19 & 1.17  & 1.08 \\ 
        ~~CRPS & 0.65 & 0.66  & 0.59 \\  
        
 \bottomrule
\end{tabular}
\end{adjustbox}
\end{table}

\end{appendices}

\end{document}